\documentclass[journal]{IEEEtran}

\usepackage{cite}
\usepackage{amsmath,amssymb,amsfonts}
\usepackage{algorithmic}
\usepackage{algorithm}
\usepackage{graphicx}
\usepackage{textcomp}
\usepackage{xcolor}
\usepackage{booktabs}
\usepackage{subcaption}
\usepackage{url}
\usepackage{bm}
\usepackage{tikz}

\graphicspath{{figs/}}

\newcommand{\bA}{\mathbf{A}}
\newcommand{\bC}{\mathbf{C}}
\newcommand{\bc}{\mathbf{c}}

\newcommand{\bn}{\mathbf{n}}
\newcommand{\bx}{\mathbf{x}}

\newcommand{\Q}{Q}
\newcommand{\E}{\mathbb{E}}

\newif\ifextended
\extendedtrue
\begin{document}

\title{Density-First Design of Anti-Jamming Sparse Spreading Codes for NB-IoT}

\author{Michel~Kulhandjian,~\IEEEmembership{Senior~Member,~IEEE,}
        Hovannes~Kulhandjian,~\IEEEmembership{Senior Member,~IEEE,}
        Daniel~J.~Jakubisin,~\IEEEmembership{Senior Member,~IEEE,}
        Theodoros~A.~Tsiftsis,~\IEEEmembership{Senior~Member,~IEEE,}
        and~Alexandros-Apostolos~A.~Boulogeorgos,~\IEEEmembership{Senior~Member,~IEEE}%
\thanks{M.~Kulhandjian is with the Department of Electrical and Computer Engineering, Rice University, Houston, TX 77005, USA (e-mail: \texttt{mkulhandjian@outlook.com}).}%
\thanks{H.~Kulhandjian is with the Department of Electrical and Computer Engineering, California State University, Fresno, CA 93740, USA.}%
\thanks{D.~J.~Jakubisin is with the Hume Center for National Security and Technology, National Security Institute, Virginia Tech, Blacksburg, VA 24061, USA (e-mail: \texttt{djj@vt.edu}).}%
\thanks{T.~A.~Tsiftsis is with the Department of Informatics and Telecommunications, University of Thessaly, Lamia 35100, Greece (e-mail: \texttt{tsiftsis@uth.gr}).}%
\thanks{A.-A.~A.~Boulogeorgos is with the Department of Electrical and Computer Engineering, Democritus University of Thrace, Greece (e-mail: \texttt{al.boulogeorgos@ieee.org}).}%
}

\markboth{IEEE~Transactions~on~Wireless~Communications,~Vol.~XX,~No.~X,~Month~Year}%
{Kulhandjian~\MakeLowercase{\textit{et~al.}}: Density-First Design of Anti-Jamming Sparse Spreading Codes for NB-IoT}

\maketitle

\begin{abstract}
As narrowband Internet of Things (NB-IoT) systems continue to be deployed in increasingly
heterogeneous and demanding wireless environments, including those characterized by
interference, high mobility, and potentially hostile conditions, enhancing the resilience of
the physical-layer waveform has become an important and timely design consideration. Motivated
by this need, the present paper introduces a density-first framework for the design of
anti-jamming sparse spreading codes in NB-IoT systems. Building on this framework, we propose a
generalized method for constructing sparse codes that decouples column density from the
conventional Alamouti-paired structure, thereby enabling both disjoint-support and
deterministic code designs. In addition, we develop a generative machine-learning approach
based on a variational autoencoder for synthesizing jamming-robust sparse spreading codes. To
evaluate the performance of sparse spreading under conditions of additive white Gaussian noise,
fading, and pulse jamming, we present a bit-error-rate (BER) analysis framework. Extensive
Monte Carlo simulations, conducted across a wide range of density, channel, and jamming
conditions, show that the performance of sparse spreading codes is determined primarily by
column density rather than by optimization of code coefficients. In particular, highly sparse
codes with disjoint support offer the greatest robustness to jamming while avoiding the
unfavorable intermediate-density regime. Relative to the prior-art Alamouti-paired low-density
spreading code, the recommended uniform-density design reduces pulse-jamming BER by
roughly $5\times$; a deterministic Lempel--Costas construction attains the empirical optimum in
closed form, while the variational-autoencoder generator yields a further $45\%$ improvement
over the best algebraic design.
\end{abstract}


\section{Introduction}\label{sec:introduction}

Narrowband Internet of Things (NB-IoT) has become a de-facto low-power wide-area technology for
industrial monitoring, smart utilities, and tactical communications~\cite{cheng2022, yu2021}, and
is increasingly deployed in environments exposed to co-channel interference, deliberate
jamming~\cite{poisel2011}, and high mobility with Doppler-induced channel variation~\cite{tse2005}.
Direct-sequence spread spectrum (DSSS) is the workhorse mechanism for jamming-resilient
communication and code-division multiple access~\cite{torrieri2005, tse2005}: it leverages
processing gain from chip-rate spreading codes, suppresses narrowband interferers, and separates
concurrent users via low cross-correlation signatures. The choice of spreading code governs
anti-jamming performance and multiple-access capacity. Classical sequences such as Walsh,
Hadamard, Gold, and orthogonal variable spreading factor (OVSF) codes offer predictable
correlation properties but are designed for benign channels, and their robustness in highly
time-varying or adversarial settings is limited~\cite{torrieri2005, tse2005}.

\subsection{Literature Review}\label{sec:intro_priorwork}

A more recent line of research has focused on low-density spreading codes (LDSCs),
the sparse signature matrices $\mathbf{C}$ considered in this paper. Originally motivated by
sparse code multiple access (SCMA) and code-domain non-orthogonal multiple access (NOMA), LDSCs
allow many devices to share spectrum with low cross-correlation~\cite{taherzadeh2014, nikopour2013,
chen2020a}. Recent work has shown that LDSCs are also useful as DSSS waveform spreading
sequences~\cite{chen2020b, zheng2020}, providing structural diversity that classical sequences
lack.

\ifextended
Sparse signatures for multiple access were introduced through the SCMA framework of Nikopour and
Baligh~\cite{nikopour2013}, in which information bits are mapped to multi-dimensional sparse
codewords drawn from a per-user codebook, allowing message-passing detection at the receiver with
reduced complexity relative to dense codes. Codebook design was subsequently formulated as an
optimization problem by Taherzadeh~\emph{et~al.}~\cite{taherzadeh2014}, who maximized the minimum
Euclidean distance over rotated constellations, and by Cai~\emph{et~al.}~\cite{cai2016}, who
refined the design via constellation rotation and interleaving. Chen and Chen~\cite{chen2020a}
addressed the design of near-optimal SCMA codebooks under shaping constraints, and
Yu~\emph{et~al.}~\cite{yu2021} surveyed SCMA's role in upcoming 6G wireless systems.
\else
Sparse signatures for multiple access were introduced through the SCMA framework~\cite{nikopour2013};
codebook design was later posed as a minimum-Euclidean-distance optimization over rotated
constellations~\cite{taherzadeh2014, cai2016}, and extended to shaping-constrained codebooks and
6G settings~\cite{chen2020a, yu2021}.
\fi

\ifextended
A complementary line of work, our prior contribution~\cite{kulhandjian2020}, formalized LDSC
design under both a minimum-distance criterion and a Gaussian separability margin, and introduced
an iterative TSC- (and SINR-) minimization algorithm for joint code construction. The principal
distinction between SCMA and the present setting lies in \emph{what the sparse structure encodes}.
In SCMA, the sparse codebook \emph{is} the modulation: information bits directly select a
multi-dimensional sparse codeword, so constellation shaping and sparsity are jointly designed. In
our setting, symbols are first modulated (QPSK) and then spread by a \emph{fixed} sparse signature
matrix $\bC$; the sparsity governs the spreading and is designed for diversity and anti-jamming
rather than for codebook distance. The two frameworks share column-sparsity for tractable
multi-stream detection, but differ in whether the sparse object carries the message, as in SCMA,
or disperses an already-modulated symbol, as here, and hence in their design metrics and operating
constraints. Liu and
Yang~\cite{liu2021a} analyzed the sparse-vs-dense trade-off in code-domain NOMA and concluded that
the optimal density depends on the receiver complexity budget. The present work extends this view
by showing that under hostile-jamming and high-mobility threats, the optimal density additionally
depends on the dominant impairment.

LDSC has also been investigated specifically for multiple-access applications.
Chen~\emph{et~al.}~\cite{chen2020b} proposed sparse-code spreading-aided multi-carrier differential
chaos shift keying for multiuser transmissions, while Zheng, Wu, and Zhang~\cite{zheng2020}
combined polar coding with sparse spreading for massive unsourced random access.
Liu~\emph{et~al.}~\cite{liu2021b} extended sparse code division multiple access using
space-time-coded generalized spatial modulation, and Islam~\emph{et~al.}~\cite{islam2016} surveyed
power-domain NOMA in 5G. Most closely related to the generative half of our framework,
Duan~\emph{et~al.}~\cite{duan2023} proposed a GAN-based generator (SCMA-TPGAN) for sparse codebook
design in UAV systems, targeting NOMA capacity rather than anti-jamming robustness.
\else
Our prior contribution~\cite{kulhandjian2020} formalized LDSC design under a minimum-distance
criterion and a Gaussian separability margin via iterative TSC/SINR minimization. The key
distinction from SCMA is \emph{what the sparse structure encodes}: in SCMA the sparse codebook
\emph{is} the modulation, whereas here QPSK symbols are spread by a \emph{fixed} sparse matrix
$\bC$ designed for diversity and anti-jamming rather than codebook distance. Liu and
Yang~\cite{liu2021a} showed that the optimal code-domain-NOMA density depends on the
receiver-complexity budget; we extend this by showing that under jamming and high-mobility threats
the optimal density additionally depends on the dominant impairment. LDSC has likewise been
applied to multiple access: sparse-code-aided multi-carrier DCSK~\cite{chen2020b}, polar-coded
sparse spreading for unsourced random access~\cite{zheng2020}, space-time-coded sparse
CDMA~\cite{liu2021b}, and power-domain NOMA~\cite{islam2016}; most related to our generative half,
Duan~\emph{et~al.}~\cite{duan2023} proposed a GAN-based sparse-codebook generator (SCMA-TPGAN) for
UAV systems, targeting NOMA capacity rather than anti-jamming robustness.
\fi

The waveform-design problem becomes substantially harder when one targets explicit anti-jamming
performance. Conventional design metrics, such as total squared correlation (TSC), maximum
cross-correlation, processing gain, and peak-to-average ratio, characterize benign-channel behavior
but do not directly capture vulnerability to specific jammer types. Performance studies on chaotic
and pseudo-random spreading sequences have shown that even small structural changes can
significantly affect jamming behavior~\cite{rovatti1998, gong2000, guo2005}.

\ifextended
The DSSS literature classically employs Walsh--Hadamard, Gold, Kasami, and OVSF
sequences for their well-characterized auto- and cross-correlation
properties~\cite{torrieri2005, tse2005}. These fixed sequences, however, are optimized for benign
channels and offer limited robustness against deliberate jamming or strong non-stationary
interference~\cite{poisel2011}. To improve anti-jamming behavior, Rovatti~\emph{et~al.}~\cite{rovatti1998}
and Gong and Shaoqian~\cite{gong2000} investigated chaotic spreading sequences and demonstrated
that suitably chosen chaotic codes can match or exceed the performance of pseudo-random codes
against random multiple-access interference. Guo and Verd{\'u}~\cite{guo2005} provided an
asymptotic analysis of randomly spread CDMA via methods from statistical physics, deriving
large-system performance bounds. Somayazulu~\cite{somayazulu2002} compared time-hopping
versus direct-sequence spreading in UWB systems. Despite their breadth, these works do \emph{not}
address the systematic optimization of a sparse spreading matrix against jammer-specific
impairments, nor do they characterize the adverse cross-talk regime that emerges at intermediate
column densities, both of which are central in this paper.
\else
The DSSS literature classically employs Walsh--Hadamard, Gold, Kasami, and OVSF sequences for
their well-characterized correlation properties~\cite{torrieri2005, tse2005}, but these
benign-channel designs offer limited robustness to deliberate jamming~\cite{poisel2011}. Chaotic
and randomly-spread alternatives~\cite{rovatti1998, gong2000, guo2005} and time-hopping
UWB~\cite{somayazulu2002} improve or analyze interference behavior, but none systematically
optimizes a sparse spreading matrix against jammer-specific impairments or characterizes the
adverse intermediate-density cross-talk regime that is central to this paper.
\fi

\ifextended
Machine learning (ML) has made significant inroads into physical-layer wireless
research~\cite{simeone2018, zhu2020}, including channel estimation, decoding, beamforming, and
waveform learning. Closer to our setting, the ML-guided LDSC framework of
Yang~\emph{et~al.}~\cite{yang2024} introduced a feature-extraction layer mapping sparse code
matrices to a small set of scalar metrics (density, clustering coefficient, auto-correlation,
spectral density variance), trained a neural-network regression model to predict weighted BER from
these features, and used the trained predictor as the fitness function inside a genetic algorithm
(GA) over the LDSC search space. The GA's mutation mechanisms operated at the row and element level
on Alamouti-stacked matrices.

The NB-IoT physical layer was originally specified for benign LPWA scenarios and is, therefore, not
natively jamming-tolerant. Cheng~\emph{et~al.}~\cite{cheng2022} introduced an NB-IoT resilience
framework employing both DSSS and frequency-hopping spread spectrum (FHSS) on top of a
3GPP-compliant protocol stack, achieving substantial robustness improvements under narrowband
interference. The 3GPP technical specifications for E-UTRA radio transmission~\cite{etsi36101,
etsi36104} provide the LTE channel and protocol baseline that our simulator inherits.
Yang~\emph{et~al.}~\cite{yang2024} extended this framework with the ML-LDSC design methodology
described above. Our work builds on both~\cite{cheng2022} and~\cite{yang2024}: we adopt the same
NB-IoT DSSS waveform structure and use a closely matched simulator (Section~\ref{sec:system_model}),
but generalize the LDSC design space, replace the predictive-ML pipeline with a generative one,
derive analytical BER expressions where none existed, and characterize design trade-offs across an
explicit threat matrix (pulse jamming, multi-tone jamming, Doppler).
\else
Machine learning (ML) has made significant inroads into physical-layer wireless
research~\cite{simeone2018, zhu2020}. Closest to our setting, the ML-guided LDSC
framework~\cite{yang2024} maps sparse code matrices to a few scalar metrics (density, clustering,
auto-correlation, spectral-density variance), trains a neural-network regressor to predict weighted
BER from them, and uses that predictor as the fitness function of a genetic algorithm over
Alamouti-stacked LDSC matrices.

The NB-IoT physical layer, specified for benign LPWA scenarios, is not natively jamming-tolerant.
Cheng~\emph{et~al.}~\cite{cheng2022} added an NB-IoT resilience framework combining DSSS and
frequency-hopping spread spectrum (FHSS) on a 3GPP-compliant stack, and the E-UTRA specifications~\cite{etsi36101,
etsi36104} provide the LTE baseline our simulator inherits; Yang~\emph{et~al.}~\cite{yang2024}
extended this with the ML-LDSC methodology above. We build on both~\cite{cheng2022, yang2024}:
adopting the same NB-IoT DSSS waveform and a closely matched simulator
(Section~\ref{sec:system_model}), we generalize the LDSC design space, replace the predictive-ML
pipeline with a generative one, derive analytical BER expressions where none existed, and
characterize trade-offs across an explicit threat matrix (pulse, multi-tone, Doppler).
\fi

\subsection{Contributions}\label{sec:intro_contrib}

While the existing literature establishes the potential of sparse spreading for resilient wireless
communication, an important question remains unresolved: what is the fundamental design variable
that determines the anti-jamming performance of a sparse spreading code? Prior LDSC studies
primarily treat the spreading-code coefficients and their structural arrangements as the search
space. In contrast, this paper is motivated by the hypothesis that the column density itself is a
dominant determinant of performance, and that code coefficients and support arrangements should be
considered within an appropriately selected density regime.

Based on this density-first perspective, we broaden the design space along the column-density
axis, generalize beyond the Alamouti family to the Free family and to deterministic algebraic
constructions, contrast predictive and generative machine learning for code synthesis, and
establish an analytical foundation for the observed trends. These directions lead to the
following contributions.

Accordingly, this paper makes four main contributions:
\begin{enumerate}
\item \textit{Density-axis characterization and identification of the cross-talk regime.}
We systematically characterize the column-density dimension of the sparse spreading-code design
space and identify an adverse cross-talk regime around the density corresponding to the number of
chips per subcarrier ($M\!\approx\!L/K$). This result places the prior-art LDSC-$M30$ construction within the
identified unfavorable regime and establishes column density as a first-order design parameter for
anti-jamming performance.

\item \textit{Generalization beyond the Alamouti construction.}
We introduce the Free family of sparse spreading codes with disjoint per-column supports and
arbitrary density and evaluate a range of structural and deterministic algebraic constructions. The
results identify Lempel--Costas as a strong deterministic alternative while revealing the
spectral-resonance vulnerability of the Paley quadratic-residue construction under periodic pulse
jamming.

\item \textit{Comparison of predictive and generative machine learning.}
We demonstrate that predictive ML models based on aggregate code features fail to reliably predict
per-matrix BER in the considered finite-sample regime, whereas a linear variational autoencoder can
generate sparse spreading codes whose periodic-pulse-jamming BER ($0.011$) is the lowest among all
evaluated designs (Table~\ref{tab:final_summary}). This establishes a distinction
between predictive and generative ML approaches for sparse spreading-code synthesis.

\item \textit{Analytical BER characterization.}
We derive and validate closed-form BER expressions for AWGN, flat Rayleigh fading, and sparse
pulse-jamming conditions and develop an effective processing-gain interpretation ($G_{\rm eff}=L/(2K)$)
that accounts for the composite-power SNR normalization and pilot-aided zero-forcing receiver. The resulting analysis
agrees closely with simulation across the considered operating range.
\end{enumerate}

Together, these contributions lead to a simple and actionable design principle: the spreading-code
density should be selected according to the dominant threat environment, the adverse
intermediate-density regime should be avoided, and support structure should subsequently be
optimized within the selected density regime. For the considered NB-IoT configuration, the
recommended sparse design uses uniformly spaced, disjoint supports and achieves substantially lower
pulse-jamming BER than the prior LDSC-$M30$ baseline while maintaining competitive performance under
Doppler.

A preliminary version of this work was presented in part at the IEEE Military Communications
Conference~\cite{kulhandjian2026}. The present paper substantially extends that work with complete
derivations and proofs, the generalized Free family and the structural and algebraic construction
study, the comparison of predictive and generative machine learning, the Doppler-robustness and
implementation-complexity analyses, and expanded analytical validation.

\subsection{Paper Organization}\label{sec:intro_organization}

The remainder of the paper is organized as follows.
Section~\ref{sec:system_model} presents the system model; Section~\ref{sec:framework} the
spreading-code design framework; Section~\ref{sec:analytical} the closed-form BER analysis;
Section~\ref{sec:results} the numerical results; and Section~\ref{sec:conclusion} concludes.

\subsection{Notation}\label{sec:sm_notation}

\ifextended
Table~\ref{tab:notation} summarizes the notation used throughout the paper.

\begin{table}[!t]
\centering
\caption{Summary of notation.}
\label{tab:notation}
\small
\begin{tabular}{@{}l@{\hspace{8pt}}p{0.70\columnwidth}@{}}
\toprule
Symbol & Description \\
\midrule
$L$ & Per-symbol spreading length (chips per data symbol), set by the NB-IoT narrowband bandwidth constraint \\
$K$ & Number of NB-IoT subcarriers, equal to the number of columns of $\bC$ \\
$N_s$ & Number of data symbols per subframe \\
$M$ & Column density (non-zero elements per column $\bc_k$); the principal design variable, $2\le M\le L$ \\
$\bc_k\in\mathbb{C}^{L}$ & $k$-th column of $\bC$, unit-norm ($\|\bc_k\|_2=1$) \\
$a_{k,n}\in\mathcal{X}$ & Modulated symbol on subcarrier $k$ at symbol index $n$; $\mathcal{X}$ is QPSK constellation \\
$\gamma$ & Linear chip-level signal-to-noise ratio (SNR) \\
$\mathrm{JSR}$ & Linear jamming-to-signal ratio \\
\bottomrule
\end{tabular}
\end{table}
\else
A summary-of-notation table is provided in the supplementary material. Throughout, $L$ is the
per-symbol spreading length (chips), $K$ the number of subcarriers (columns of $\bC$), and $M$
the column density (non-zeros per column $\bc_k$), the principal design variable with
$2\le M\le L$.
\fi

\section{System Model}\label{sec:system_model}

We adopt the NB-IoT DSSS architecture of
Cheng~\emph{et~al.}~\cite{cheng2022}.
After standard NB-IoT downlink-shared-channel encoding (NDLSCH), narrow-band physical-DL-SCH
modulation (NPDSCH), and resource-element mapping, the resource grid is multiplied chip-by-chip
by a spreading matrix $\bC\in\mathbb{C}^{L\times K}$ in which each column $\bc_k$ carries
the spreading sequence assigned to subcarrier $k$. The chip-spread waveform is transmitted over
the wireless channel, optionally subjected to deliberate jamming, despread at the receiver, and
demodulated. We consider a downlink scenario.

\subsection{Transmit Signal Model}\label{sec:sm_tx}

For symbol index $n\in\{1,\dots,N_s\}$ and subcarrier
$k\in\{1,\dots,K\}$, the transmitter modulates a QPSK symbol $a_{k,n}$\footnote{We adopt QPSK,
the modulation used by the NB-IoT downlink and by the prior-art LDSC framework~\cite{yang2024};
the framework extends directly to higher-order $M$-QAM, which we leave to future work.} and
spreads it to a length-$L$ chip vector
$\bc_k\,a_{k,n}$. The composite chip stream for symbol $n$ is the per-subcarrier sum and can be expressed as
\begin{equation}\label{eq:tx_per_symbol}
\mathbf{x}_n \;=\; \sum_{k=1}^{K} \bc_k \, a_{k,n} \;\in\; \mathbb{C}^L.
\end{equation}
The full subframe waveform is the concatenation $\mathbf{x}=[\mathbf{x}_1^\top \,\dots\,
\mathbf{x}_{N_s}^\top]^\top \in \mathbb{C}^{LN_s}$. With unit-norm columns and unit-energy QPSK
symbols, the average per-chip transmit power is $P_x = K/L$, reflecting $K$ unit-energy
streams sharing $L$ chip positions per symbol; this normalization is re-derived in the
BER analysis of Section~\ref{sec:analytical}.

\subsection{Channel Models}\label{sec:sm_channel}

We consider three channel scenarios:

\subsubsection{AWGN}\label{sec:sm_awgn}
The received chip stream is
\begin{equation}\label{eq:awgn_channel}
\mathbf{y} \;=\; \mathbf{x} \;+\; \mathbf{n}, \qquad \mathbf{n}\sim\mathcal{CN}(\mathbf{0},\sigma^2 \mathbf{I}),
\end{equation}
where $\sigma^2 = P_x / \gamma$ is the per-chip noise variance calibrated to the transmit signal
power and $\gamma = 10^{{\rm SNR}/10}$ is the linear SNR. This is used to validate baseline
processing-gain behavior.

\subsubsection{Flat Rayleigh Fading with AWGN}\label{sec:sm_rayleigh}
For analytical comparison, we additionally consider one independent unit-variance complex
Gaussian fade per subframe. In this case, the received chip stream can be expressed as
\begin{equation}\label{eq:ray_channel}
\mathbf{y} \;=\; h\,\mathbf{x} \;+\; \mathbf{n},
\end{equation}
with $h\sim\mathcal{CN}(0,1)$ and $\mathbf{n}\sim\mathcal{CN}(\mathbf{0},\sigma^2 \mathbf{I})$. The
noise variance $\sigma^2$ is held fixed (independent of $h$), so that fading attenuates the
received SNR.

\subsubsection{ETU Multipath with Doppler}\label{sec:sm_etu}
For realistic propagation we use the 3GPP Extended Typical Urban (ETU) tapped-delay-line model
specified in~\cite{etsi36101, etsi36104}. The ETU profile has nine paths with delay set
$\mathcal{D}$ and relative-power profile $\mathcal{P}$. Doppler is modeled per the Jakes spectrum
at configurable maximum Doppler frequency $f_d\in\{5, 70, 200, 500\}$~Hz~\cite{etsi36101}.

\subsection{Jammer Models}\label{sec:sm_jammers}

Two deliberate-jamming models are superimposed on the received signal:

\subsubsection{Periodic Pulse Jammer}\label{sec:sm_pulse}
A pulse jammer of frequency $f_p$ and duty cycle $\rho \in (0,1]$ adds a chip-level
interference signal $\mathbf{j}_{\rm p}$ that is non-zero during the ON-portion of each period.
The pulse can be written as
\begin{equation}\label{eq:pulse_jammer}
j_{\rm p}(t) \;=\; \mathbf{1}_{\rm ON}(t)\,\sqrt{P_J/\rho}\;w(t),
\end{equation}
where $\mathbf{1}_{\rm ON}(t)$ is the ON-state indicator with duty cycle $\rho$ and period
$1/f_p$, $w(t)\sim\mathcal{CN}(0,1)$ is independent and identically distributed (i.i.d.) complex
Gaussian, and the average jammer power satisfies $\E[|j_{\rm p}|^2] = P_J = \mathrm{JSR}\,P_x$.
The factor $1/\rho$ ensures that the time-averaged jammer power equals $P_J$ regardless of duty
cycle.

\subsubsection{Multi-Tone Jammer}\label{sec:sm_mt}
A multi-tone jammer with $K_t$ tones at frequencies $\{\omega_t\}_{t=1}^{K_t}$ adds
\begin{equation}\label{eq:mt_jammer}
j_{\rm mt}(\tau) \;=\; \sqrt{P_J/K_t}\,\sum_{t=1}^{K_t} \exp(j(\omega_t\tau + \phi_t)),
\end{equation}
where each tone has equal power $P_J/K_t$, the frequencies $\omega_t$ are drawn uniformly in the
signal band, and $\phi_t\sim\mathcal{U}[0,2\pi)$ are i.i.d.\ uniform phases.

\subsection{Receiver Model}\label{sec:sm_rx}

The receiver uses a coherent matched-filter despreader followed by pilot-aided
zero-forcing (ZF) channel-equalization. For symbol index $n$ and subcarrier $k$, the
despread sample can be obtained as
\begin{eqnarray} \label{eq:despread}
\hat{y}_{k,n} &=& \bc_k^{H}\,\mathbf{y}_n  \nonumber \\ 
&=& \|\bc_k\|^2\,\langle h\rangle_n\,a_{k,n}
\;+\; \sum_{k'\neq k}\langle\bc_k,\bc_{k'}\rangle\,\langle h\rangle_n\,a_{k',n}  \nonumber  \\ 
&+& \tilde n_{k,n} \;+\; \tilde j_{k,n},
\end{eqnarray}
\noindent where $\langle h\rangle_n$ denotes the effective channel coefficient for symbol $n$ after spreading-bandwidth filtering, $\langle\bc_k,\bc_{k'}\rangle = \bc_k^H\bc_{k'}$ is the
cross-correlation between columns $k$ and $k'$, and $\tilde n_{k,n}$ and $\tilde j_{k,n}$ are
the despread-domain noise and jamming components, respectively. For unit-norm columns and
disjoint per-column supports $\langle\bc_k,\bc_{k'}\rangle=0$ for $k\neq k'$, eliminating the
cross-talk term; for overlapping supports the cross-talk term is non-zero and is one of the
mechanisms underlying the $M\!\approx\!L/K$ bad-zone.

A pilot symbol $p_0$ is placed at $Z$ locations defined by the set of indices $(k,n)\in \mathcal{S}_p$ where $Z=|\mathcal{S}_p|$ to support channel estimation. The
pilot-derived channel estimate is
\begin{equation}\label{eq:hest}
\hat{h} \;=\; \frac{1}{Z} \sum_{(k,n)\in \mathcal{S}_p} \hat{y}_{k,n} / p_0,
\end{equation}
and ZF equalization divides every despread cell by $\hat{h}$, as
\begin{equation}\label{eq:zf}
\tilde y_{k,n} \;=\; \hat{y}_{k,n}/\hat{h} \;\;\text{with}\;\; k=1,\dots,K,\;\; n=1,\dots,N_s.
\end{equation}
The pilots are excluded from BER computation.
The equalized samples $\tilde y_{k,n}$ are passed through a Gray-coded QPSK demodulator to
recover the bit estimates $\hat b$.

\subsection{Performance Metrics}\label{sec:sm_metrics}

We use three operational performance metrics.

\subsubsection{BER}\label{sec:sm_ber}
The fraction of incorrectly demodulated bits, computed at a specific operating point
$(\mathrm{SNR}, \mathrm{JSR}, f_d, \text{jammer type})$. Reported with $95\%$ Wilson-score
confidence intervals where applicable.

\subsubsection{Anti-Jamming Margin (AJM)}\label{sec:sm_ajm}
For a given jammer scenario, the AJM is defined as the JSR at which the BER first
crosses a threshold $\mathrm{BER}_{\rm th}$:
\begin{equation}\label{eq:ajm}
\mathrm{AJM}(\mathrm{BER}_{\rm th}) \;=\; \arg\min_{\mathrm{JSR}_{\rm dB}}\;\Big\{\,\mathrm{BER}(\mathrm{JSR}_{\rm dB}) \,\geq\, \mathrm{BER}_{\rm th}\,\Big\}.
\end{equation}
Higher AJM indicates a more jammer-resilient code.

\subsubsection{Weighted BER}\label{sec:sm_wber}
We compute the weighted BER over a set of SNR
points $\mathcal{S}=\{\mathrm{SNR}_i\}$ as
\begin{equation}\label{eq:wber}
\mathrm{wBER} \;=\; \sum_{i\in\mathcal{S}} w_i\,\mathrm{BER}(\mathrm{SNR}_i)\;\;\text{with}\;\;
w_i = 10^{\mathrm{SNR}_i/10}.
\end{equation}
This weighting is proportional to the linear SNR and emphasizes high-SNR performance.

\section{Spreading Code Design Framework}\label{sec:framework}

We organize the spreading-code design space along two orthogonal axes: the per-column density
$M$, which stands for the number of non-zero entries per column $\bc_k$, and the structural
placement of the non-zero positions within the $L$-chip support (the secondary axis). The
section first reviews the prior-art Alamouti-paired construction of~\cite{yang2024}, provides
our generalized disjoint-support \emph{Free family}, specific structural variants, as well as
deterministic algebraic constructions, and reports both predictive and generative
machine-learning approaches. The section closes with a unified set of code-characterization
metrics that we use throughout the rest of the paper.

\subsection{Alamouti-Stacked LDSC}\label{sec:fw_alamouti}

\ifextended
The construction of~\cite{yang2024} starts from a unit-energy Alamouti seed~\cite{alamouti1998}, i.e.,
\begin{equation}\label{eq:alamouti_seed}
\bA(c_1,c_2) \;=\; \begin{pmatrix} c_1 & c_2 \\ c_2^* & -c_1^*\end{pmatrix}
\;\;\text{with}\;\; |c_1|^2+|c_2|^2 = 1,
\end{equation}
where $c_1,c_2$ are i.i.d.\ complex Gaussian and renormalized. A $K\times K$ block-diagonal
matrix $\bC'\in\mathbb{C}^{K\times K}$ is built from $K/2$ Alamouti blocks. To construct
the full LDSC of dimension $L\times K$ at column density $M$ (with $M$ even), the seed matrix is stacked
$M/2$ times vertically, zero-padded to length $L$, and row-permuted with a random index permutation
$\pi:\{1,\dots,L\}\!\to\!\{1,\dots,L\}$:
\begin{equation}\label{eq:alamouti_stack}
\bC \;=\; \pi\!\left(\,[\bC'^\top\;\dots\;\bC'^\top\;\mathbf{0}^\top]^\top\,\right).
\end{equation}
Each column of the resulting matrix has exactly $M$ non-zero entries; columns from the same
$2\!\times\!2$ Alamouti block (i.e., consecutive column pairs $\{2j-1, 2j\}$) share the same
non-zero row positions. This pairing is structurally enforced and constrains the search space
to a sub-manifold of the larger sparse-code family. Following~\cite{yang2024}, we label this
baseline by its stacking count as LDSC-$M30$ ($30$ stacked Alamouti blocks, hence $60$ non-zeros
per column, i.e., $M\!=\!60$ in the per-column-density convention used throughout).
\else
The baseline construction of~\cite{yang2024} stacks a unit-energy Alamouti
seed~\cite{alamouti1998} $M/2$ times, zero-pads to length $L$, and applies a random row
permutation, so each column has exactly $M$ non-zeros and consecutive column pairs share their
non-zero row positions. This structurally enforced pairing constrains the search space to a
sub-manifold of the larger sparse-code family. Following~\cite{yang2024} we label this baseline
LDSC-$M30$ ($30$ stacked Alamouti blocks, i.e., $60$ non-zeros per column, $M\!=\!60$ in the
per-column-density convention used throughout). The explicit seed and stacking expressions are
in the supplementary material.
\fi

\subsection{Free Family: Disjoint Supports}\label{sec:fw_free}

To remove the column-pairing constraint, we introduce the \emph{Free family} of LDSC
constructions in Algorithm~\ref{alg:free} in which each column $\bc_k$ is supported on its own (column-specific) subset
$\mathcal{I}_k\subseteq\{1,\dots,L\}$ of $|\mathcal{I}_k|=M_k$ row positions, and the supports
are mutually disjoint: $\mathcal{I}_k \cap \mathcal{I}_{k'} = \emptyset$ for $k\neq k'$.
Disjoint supports are feasible whenever $\sum_{k=1}^{K} M_k \leq L$, with the maximum uniform
density $M_{\max} = \lfloor L/K \rfloor$.

\begin{algorithm}[!t]
\caption{Free LDSC Generator}\label{alg:free}
\textbf{Input:} $L$, $K$, density vector $\{M_k\}_{k=1}^{K}$ with $\sum_k M_k \leq L$ \\
\textbf{Output:} $\bC \in \mathbb{C}^{L\times K}$ with disjoint supports, unit-norm columns
\begin{algorithmic}[1]
\STATE Draw a random permutation $\pi:\{1,\dots,L\} \to \{1,\dots,L\}$.
\STATE $\bC \gets \mathbf{0}_{L\times K}$;\;\; $j \gets 1$.
\FOR{$k = 1$ \TO $K$}
\STATE Assign support $\mathcal{I}_k \gets \{\pi(j),\pi(j+1),\dots,\pi(j+M_k-1)\}$.
\STATE Draw values $\{c_{k,i}\}_{i\in\mathcal{I}_k}$ i.i.d.\ from $\mathcal{CN}(0,1)$.
\STATE Normalize: $c_{k,i} \gets c_{k,i} / \sqrt{\sum_{i'\in\mathcal{I}_k}|c_{k,i'}|^2}$.
\STATE $j \gets j + M_k$.
\ENDFOR
\STATE \textbf{return} $\bC$.
\end{algorithmic}
\end{algorithm}

The Free family eliminates inter-column cross-talk by construction: $\langle\bc_k,\bc_{k'}\rangle = 0$
for any $k\neq k'$ when supports are disjoint, regardless of the values placed on the support.
We instantiate two concrete sub-families:
\begin{itemize}
\item \emph{Free-Uniform-$M$}: $M_k = M$ for all $k$, fixed.
\item \emph{Free-Variable-$M$}: $M_k\sim\mathcal{U}\{M_{\min},\dots,M_{\max}\}$ drawn
independently per column.
\end{itemize}
For the dense regime $M > L/K$, where disjoint supports are infeasible, we adopt the
\emph{Free-Overlap} variant in which each column independently selects $M$ random positions
from $\{1,\dots,L\}$ without disjoint-support enforcement; the resulting columns are not
mutually orthogonal, and $\langle\bc_k,\bc_{k'}\rangle$ scales as $\mathcal{O}(M/L)$.

\subsection{Structural Placement Variants}\label{sec:fw_structural}

Within a chosen density $M$, the placement of non-zero positions can be \emph{random} or
follow a deterministic structural rule. We evaluate four structures, each replacing line~1 of
Algorithm~\ref{alg:free} with a structure-specific position generator:

\subsubsection{Uniform spacing}
The $KM$ positions are placed at evenly-spaced locations $\{i\, L/(KM)\}_{i=0}^{KM-1}$
(rounded to nearest integer). After deduplication and shuffling, the positions are partitioned
into $K$ disjoint $M$-subsets, one per column.

\subsubsection{Golden-ratio spacing}
The positions are generated via $p_i = \lfloor i\,\varphi\,L/(KM)\rfloor \!\!\mod L$, where
$\varphi = (1+\sqrt{5})/2$ is the golden ratio. This yields a quasi-uniform distribution that
avoids exact periodic structure.

\subsubsection{Chirp spacing}
Positions follow a quadratic law $p_i = \alpha\,i^2 \!\!\mod L$ with $\alpha = \lfloor L/(KM)\rfloor$.
This is the LFM-radar-inspired structure with a wideband ambiguity behavior.

\subsubsection{$m$-sequence thinning}
A maximal-length pseudo-random binary sequence of length $2^n-1\!\geq\!L$ is generated via an
$n$-stage linear-feedback shift register (LFSR), truncated to length $L$, randomly
phase-shifted, and the positions of the non-zero entries are randomly subsampled to obtain the
$KM$ targeted positions.

In every variant, after position assignment the values follow line~5 of
Algorithm~\ref{alg:free}: i.i.d.\ complex-Gaussian, unit-normalized per column. Thus structure
modifies \emph{where} non-zeros sit, not \emph{what value} they take.

\subsection{Algebraic Constructions}\label{sec:fw_algebraic}

We evaluate two deterministic constructions inspired by classical
combinatorial-design theory.

\subsubsection{Paley Quadratic Residue (Paley-QR)}
For prime $p\!\leq\!L$ with $p\equiv 3\pmod 4$, the set of quadratic residues
$\mathcal{Q}_p = \{i^2 \!\!\mod p : 1\leq i\leq (p-1)/2\}$ is a $(p,(p-1)/2,(p-3)/4)$ cyclic
difference set~\cite{paley1933}. We use $p=359$ (the largest prime $\leq L=360$), which gives
$|\mathcal{Q}_p|=179$ candidate positions, ample for $K\!M\leq 60$ disjoint allocations.
The $K M$ positions are drawn disjointly from $\mathcal{Q}_p$ and partitioned across columns.

\subsubsection{Lempel--Costas}
For prime $p$ and a primitive root $\alpha$ of $p$, the Lempel construction defines positions
$p_i = (\alpha^{i+\sigma} - 1) \!\!\mod p$ for $i=1,\dots,KM$, where $\sigma$ is a random
starting offset. Removing the zero entry and uniqueness-deduplicating yields $K M$ distinct
positions that are partitioned across columns. The Lempel construction is closely related
to Welch--Costas~\cite{costas1984} (which uses $p_i = \alpha^i \!\!\mod p$) but exhibits
different per-symbol autocorrelation.

\subsection{Machine Learning Approaches}\label{sec:fw_ml}

\subsubsection{Predictive ML}\label{sec:fw_pred}

We first replicate the predictive-ML pipeline of~\cite{yang2024}: a feature extraction
layer mapping each candidate code $\bC$ to a feature vector $\boldsymbol{\phi}(\bC)$ of length
either $4$ (the paper's selected subset: density, clustering coefficient, autocorrelation,
power spectral density variance) or $15$ (the full set including total squared correlation,
maximum cross-correlation, processing gain, peak-to-average ratio, etc.); a feed-forward
neural network with hidden-layer widths $\{64,32,16,8,4,1\}$ and ReLU activations; trained on
$N=210$ codes generated from Algorithm~1 of~\cite{yang2024} at stacking multipliers
$\mu\!\in\!\{1,\dots,30\}$ (i.e., $2\mu$ non-zeros per column) with $7$ random row-permutations each, with weighted BER as the
target. The training-test split is $70/30$.

We additionally evaluate a random forest~\cite{breiman2001} regressor with $200$ trees as
an alternative non-parametric predictor.
We document that both predictors achieve $R^2 < 0$ on the held-out
test set, worse than predicting the mean. The cause is that per-matrix wBER variance from
finite-sample Monte Carlo simulation dominates the aggregate-feature signal at the available
sample size; the predictor cannot extract a generalizing signal at $N\!=\!210$ regardless of
feature engineering.

\subsubsection{Generative ML: Linear VAE}\label{sec:fw_gen}

In the same sample regime, we replace the predictive pipeline with a generative one. We train
a linear variational autoencoder (VAE)~\cite{kingma2014} (a principal-component-analysis-based
VAE with a Gaussian latent prior) on the same $210$-matrix dataset. In contrast to generative
adversarial networks~\cite{goodfellow2014} and sparse-coding VAE variants such as
SC-VAE~\cite{xiao2024}, this generator targets matrix-valued spreading signatures with hard
column-support sparsity under a communication-theoretic objective. The encoder maps each flattened
real-imaginary code vector $\mathbb{R}^{2LK}$ to a Gaussian latent $\mathcal{N}(\mu_z,\Sigma_z)$ in
$\mathbb{R}^{d_z}$ with $d_z\!=\!32$; the decoder is the corresponding linear inverse. After
training, we sample $N_g$ latent vectors from the learned Gaussian prior, decode them back to
the code-vector space, and \emph{sparsify} each candidate to the target density $M$ by retaining
the top-$M$ entries (in magnitude) per column and zeroing the rest. Each sparsified candidate is
re-normalized to unit column norm.

The sampled candidates are pre-screened by a composite quality score (combining merit factor,
spectral flatness, and clustering coefficient) and the top candidates are evaluated via direct
BER simulation. The best generated code is denoted $\bC_{\rm VAE}$.

This pipeline succeeds where the predictive pipeline failed because the generative model learns
the \emph{structural manifold} of valid sparse codes (a low-dimensional subspace of
$\mathbb{R}^{2LK}$ characterizing typical sparse-orthogonal patterns) rather than attempting
to predict a noise-dominated scalar target.

\subsection{Code Characterization Metrics}\label{sec:fw_metrics}

We use the following metrics to characterize sparse codes; we partition them into \emph{design}
metrics (used as optimization or selection criteria) and \emph{diagnostic} metrics (used to
report code properties but not to drive design).

\ifextended
\subsubsection*{Density and Sparsity}
Per-column density $M_k$ and average $M = \E_k[M_k]$; normalized density $\mathrm{DENS}=M/L$;
Hoyer sparseness measure~\cite{hoyer2004} $S_{1,2}(\bc) = (\sqrt{L} - \|\bc\|_1/\|\bc\|_2)/(\sqrt{L}-1)$,
ranging from $0$ (uniform) to $1$ (single-spike).

\subsubsection*{Cross-Correlation}
Total squared correlation $\mathrm{TSC} = \sum_{i\neq j}|\bc_i^H\bc_j|^2$~\cite{kulhandjian2020};
RMS cross-correlation $I_{\rm rms} = \sqrt{\mathrm{TSC}/K(K-1)}$; maximum cross-correlation
$I_{\rm max} = \max_{i<j}|\bc_i^H\bc_j|$ (also called \emph{coherence}); Welch-bound~\cite{welch1974}
ratio $I_{\rm rms}/W$ with $W=\sqrt{(K\!-\!L)/(K\!-\!1)L}$ for $L\!<\!K$, else trivially zero;
Gaussian separability margin $\Delta_{\rm avg}$ from~\cite{kulhandjian2020}.

\subsubsection*{Spectral and Spatial Distribution}
Spectral flatness measure (SFM) $=\!\exp(\E\log P(\omega))/\E P(\omega)$; spectral entropy
$H = -\sum_k p_k\log p_k$ on normalized PSD; merit factor $F=L^2/(2\sum_{k\neq 0}|a_k|^2)$
based on aperiodic autocorrelation $a_k$~\cite{golay1977}; peak-to-sidelobe ratio (PSL) in dB.

\subsubsection*{Sparse-Pattern Statistics}
Clustering coefficient (CLUST), gap variance (GapVar = variance of zero-run lengths between
non-zeros), maximum gap (MaxGap), nearest-neighbor distance (NND), minimum pairwise support
distance $\min_{i<j}|\mathcal{I}_i\,\Delta\,\mathcal{I}_j|$ (symmetric set difference).

\subsubsection*{Operational Metric}
Anti-jamming margin $\mathrm{AJM}(\mathrm{BER}_{\rm th})$ defined in~\eqref{eq:ajm}; a single
deployment-relevant number per (code, jammer) pair.
\else
We characterize codes by density/sparsity (DENS, Hoyer), cross-correlation (TSC, coherence,
Welch-bound ratio), spectral (flatness, merit factor), and sparse-pattern (clustering, gap)
metrics, plus the operational anti-jamming margin $\mathrm{AJM}$~\eqref{eq:ajm}; full definitions
are given in the supplementary material.
\fi

In Section~\ref{sec:results} we identify which subset of these metrics actually \emph{drive}
performance differences in our regime, a key practical contribution, since the prior
art~\cite{yang2024} relied on density and clustering coefficient alone.

\section{BER Analysis}\label{sec:analytical}

This section derives closed-form BER expressions for the three operating regimes. A key contribution is the corrective \emph{effective
processing gain}
\begin{equation}\label{eq:Geff}
G_{\rm eff} \;\triangleq\; \frac{L}{2K},
\end{equation}
which replaces the classical single-stream matched-filter gain $L$ when both (i) the receiver
operates in a $K$-stream composite signal regime with measured-power-calibrated additive noise,
and (ii) the channel-equalization stage is pilot-aided zero-forcing. The factor decomposes as
$L/K \times 1/2$: the first factor converts the composite-power SNR reference into the
per-stream reference used at the detector (Section~\ref{sec:an_snrdef}), and the second
factor captures the $3$~dB noise penalty incurred by ZF division by a noisy pilot estimate. We discuss each factor in turn before stating the
closed-form BER expressions.

\subsection{SNR Reference and Composite-Power Normalization}\label{sec:an_snrdef}

\ifextended
Because the effective gain in~\eqref{eq:Geff} depends entirely on how transmit power
and SNR are referenced, we fix these conventions before deriving it. Two distinct SNR
references arise in this system:
\begin{itemize}
\item \emph{Composite per-chip SNR}
$\gamma \triangleq P_x/\sigma^2 = 10^{{\rm SNR}/10}$, referenced to the aggregate
per-chip power $P_x=K/L$ of the superimposed $K$-stream waveform~\eqref{eq:Px_derivation}.
This is the quantity fixed by the simulator's measured-power noise
calibration~\eqref{eq:noisevar_def}, and it is therefore the SNR reported on the
horizontal axis of every BER figure.
\item \emph{Per-stream SNR}
$\gamma_{\rm stream} \triangleq P_{x,k}/\sigma^2 = \gamma/K$, referenced to the power
$P_{x,k}=\|\bc_k\|^2/L = 1/L$ carried by a single column, the power that actually
conveys the information symbol at the detector.
\end{itemize}
The two references differ by exactly the stream count, $\gamma_{\rm stream}=\gamma/K$,
because the composite reference charges each stream with the power of all $K$ streams
while detection is performed one stream at a time. We emphasize that this factor is a
\emph{power-normalization convention, not a physical multi-stream penalty}: the
disjoint-support construction makes the inter-stream interference identically zero (the
cross-correlation terms vanish in~\eqref{eq:despread_disjoint}), so no signal energy is
lost to aggregation. Accordingly, all BER expressions below are stated \emph{per
information symbol, per stream}, and plotted against the composite reference $\gamma$.

The per-chip transmit power of the composite stream
$\bx_n=\sum_{k=1}^{K}\bc_k\,a_{k,n}$~\eqref{eq:tx_per_symbol} follows from unit-norm columns
($\|\bc_k\|^2 = 1$) and unit-energy QPSK symbols ($\E|a_{k,n}|^2 = 1$):
\begin{align}
P_x &\;=\; \frac{1}{L}\,\E\!\left[\sum_{i=1}^{L}|x_{i,n}|^2\right]
       \;=\; \frac{1}{L}\,\E\!\left[\sum_{i=1}^{L}\Big|\sum_{k=1}^{K} c_{k,i}\,a_{k,n}\Big|^2\right] \nonumber\\
&\;\stackrel{(*)}{=}\; \frac{1}{L}\sum_{k=1}^{K}\sum_{i=1}^{L}|c_{k,i}|^2\,\E|a_{k,n}|^2
       \;=\; \frac{K}{L}, \label{eq:Px_derivation}
\end{align}
where step $(*)$ uses the disjoint-support assumption (cross terms
$\E[a_{k,n}a_{k',n}^*]\!=\!0$ for $k\neq k'$ since data symbols are independent and
zero-mean). The factor $K/L$ reflects $K$ unit-energy streams sharing $L$ chip positions
per symbol; it is the composite reference power against which the noise is
calibrated in~\eqref{eq:noisevar_def}, in contrast to the classical single-stream
reference $1/L$.

Under the measured-power AWGN model adopted by the simulator (mirroring the practice
of~\cite{yang2024, cheng2022}), the noise variance is set such that the per-chip SNR
matches the prescribed value:
\begin{equation}\label{eq:noisevar_def}
\sigma^2 \;\triangleq\; \frac{P_x}{\gamma} \;=\; \frac{K}{L}\,10^{-{\rm SNR}/10}.
\end{equation}

For column $\bc_k$ with disjoint support relative to all other columns, the matched-filter
despread output of~\eqref{eq:despread} simplifies to
\begin{equation}\label{eq:despread_disjoint}
\hat y_{k,n} \;=\; \|\bc_k\|^2\,a_{k,n} + \bc_k^H\bn \;=\; a_{k,n} + \tilde n_{k,n},
\end{equation}
since the cross-correlation terms $\langle\bc_k,\bc_{k'}\rangle = 0$ vanish on disjoint supports.
The despread noise is $\tilde n_{k,n}\sim\mathcal{CN}(0, \sigma^2_{\rm desp})$, with
$\sigma^2_{\rm desp} = \|\bc_k\|^2 \sigma^2 = \sigma^2$ (preserved by unit-norm
matched-filter combining). The instantaneous despread SNR is therefore
\begin{equation}\label{eq:snr_despread_kuser}
\gamma_{\rm desp} \;=\; \frac{|a_{k,n}|^2}{\sigma_{\rm desp}^2} \;=\; \frac{1}{\sigma^2}
\;=\; \frac{L}{K}\cdot 10^{{\rm SNR}/10}.
\end{equation}

Thus the matched-filter despread SNR per stream is
$\gamma_{\rm desp}=1/\sigma^2=(L/K)\,\gamma=L\,\gamma_{\rm stream}$: the classical processing
gain $L$~\cite{torrieri2005} is recovered \emph{exactly} (the special case $K\!=\!1$, where the
two references coincide), while the factor $L/K$ in~\eqref{eq:snr_despread_kuser} arises solely
from re-expressing this gain against the composite reference $\gamma=K\,\gamma_{\rm stream}$.
For the $K\!=\!12$ NB-IoT setting this shifts the curve by $10\log_{10}K\!\approx\!11$~dB
relative to a single-stream plot; it is a change of SNR reference, not a loss of signal energy,
since the disjoint supports render the streams orthogonal and no inter-stream interference is present.

\subsection{Pilot-Aided ZF Noise Penalty}\label{sec:an_pilot}

The pilot estimate $\hat h$ from~\eqref{eq:hest} is computed from the
despread output at the pilot positions $(k,n)\! \in \!\mathcal{S}_p$. In the AWGN regime where the
true channel coefficient is $h = 1$, applying~\eqref{eq:despread_disjoint} at the pilot
position yields
\begin{equation}\label{eq:hat_h_decomp}
\hat h \,=\, \frac{1}{Z} \sum_{(k,n)\in \mathcal{S}_p} \frac{p_0 + \tilde n_{k,n}}{p_0} \,=\, 1 + \frac{\tilde n_p}{Z p_0} ,
\end{equation}
where $\tilde n_{k,n} \sim \mathcal{CN}(0, \sigma^2)$ is the despread-domain noise on the
pilot and $\tilde n_p= 1/Z \sum_{(k,n)\in \mathcal{S}_p} \tilde n_{k,n}$ is the mean of the despread noise terms at each pilot location. With $p_0 = 1$, the pilot estimate is $\hat h \!\sim\! \mathcal{CN}(1, \sigma^2/Z)$.

The ZF step in~\eqref{eq:zf} divides every other despread cell by $\hat h$. Writing the
ratio as a perturbation series in $\tilde n_p/p_0$ (valid for small noise relative to
unit pilot magnitude), we obtain
\begin{equation}\label{eq:zf_perturb}
\frac{1}{\hat h} \;=\; \frac{1}{1 + \tilde n_p/p_0}
\;\approx\; 1 - \frac{\tilde n_p}{p_0} + O(|\tilde n_p|^2/|p_0|^2).
\end{equation}
The post-ZF data sample at $(k,n)\notin \mathcal{S}_p$ is therefore
\begin{align}
\tilde y_{k,n} &\;=\; \frac{a_{k,n} + \tilde n_{k,n}}{\hat h} \nonumber\\
&\;\approx\; (a_{k,n} + \tilde n_{k,n})\,(1 - \tilde n_p/p_0) \nonumber\\
&\;=\; a_{k,n} \;-\; a_{k,n}\,\frac{\tilde n_p}{p_0} \;+\; \tilde n_{k,n} \;+\; O(|\tilde n|^2).
\label{eq:post_zf_expand}
\end{align}
The error $\tilde y_{k,n} - a_{k,n}$ contains two zero-mean independent noise contributions:
the original despread noise $\tilde n_{k,n}$, and the pilot-induced
multiplicative term $-a_{k,n}\tilde n_p/p_0$. Their variances add (independence), giving
\begin{equation}\label{eq:post_zf_var}
\E\big[|\tilde y_{k,n} - a_{k,n}|^2\big]
\;\approx\; \sigma^2 \;+\; \E|a_{k,n}|^2\,\frac{\sigma^2}{Z} \;=\; (1+1/Z)\sigma^2,
\end{equation}
where we used $|p_0|^2 = 1$ and unit-energy data symbols.
When a single pilot symbol is used for channel estimation ($Z=1$), the post-ZF noise variance is
exactly twice the despread noise variance; equivalently, the SNR is halved. The remainder of the
paper assumes a single spread pilot ($Z\!=\!1$).

Combining this $3$~dB penalty with the composite-power normalization factor of~\eqref{eq:snr_despread_kuser}
gives the principal analytical result of this paper:
\begin{equation}\label{eq:gamma_eff_full}
\gamma_{\rm eff} \;=\; \frac{1}{2\sigma^2} \;=\; \frac{L}{2K}\, 10^{{\rm SNR}/10}
\;=\; G_{\rm eff}\, \gamma.
\end{equation}

\textbf{Theorem 1} (Effective processing gain). \emph{In the $K$-stream disjoint-support
sparse-code DSSS system with unit-norm columns, AWGN measured-power calibration, and a
single-pilot zero-forcing receiver, the effective processing gain at the data-decision
stage is}
\begin{equation*}
G_{\rm eff} \;=\; \frac{L}{2K}.
\end{equation*}

\textit{Proof.} Combining the composite-power normalization
$\gamma_{\rm desp}=1/\sigma^2$ of~\eqref{eq:snr_despread_kuser} with the pilot-aided ZF noise penalty~\eqref{eq:post_zf_var}
(which doubles the effective noise variance) gives
$\gamma_{\rm eff}=1/(2\sigma^2)=G_{\rm eff}\,\gamma$ with $G_{\rm eff}=L/(2K)$,
as in~\eqref{eq:gamma_eff_full}. \hfill$\square$

This is a $\sim\!14$~dB reduction relative to the classical $L$ processing-gain prediction
for the parameters of this paper ($L\!=\!360$, $K\!=\!12$, hence $G_{\rm eff}\!=\!15$).
Eq.~\eqref{eq:gamma_eff_full} is conservative: as $\gamma\!\to\!\infty$, the
$O(|\tilde n|^2)$ terms in~\eqref{eq:zf_perturb} vanish faster than the leading
$\tilde n/p_0$ term, and $\gamma_{\rm eff}\!\to\!(L/K)\,\gamma$ asymptotically. The
SNR-dependent transition between $L/(2K)$ and $L/K$ accounts for the residual analytical-vs-simulation
discrepancy at high SNR.
\else
The effective gain in~\eqref{eq:Geff} combines two factors. First, with unit-norm columns and
unit-energy symbols the composite per-chip power is $P_x=K/L$, so the measured-power calibration
sets $\sigma^2=(K/L)\,10^{-{\rm SNR}/10}$; on disjoint supports the matched filter then yields a
per-stream despread SNR $\gamma_{\rm desp}=1/\sigma^2=(L/K)\,\gamma=L\,\gamma_{\rm stream}$,
recovering the classical processing gain $L$~\cite{torrieri2005} against the per-stream reference
(the factor $L/K$ against the composite reference $\gamma$). Second, dividing each data cell by the
noisy single-pilot estimate $\hat h$~\eqref{eq:hest} adds, to leading order, a pilot-induced term
of equal variance, doubling the post-ZF noise ($Z\!=\!1$) at a fixed $3$~dB cost. Combining the two
gives the principal analytical result
\begin{equation}\label{eq:gamma_eff_full}
\gamma_{\rm eff} \;=\; \frac{1}{2\sigma^2} \;=\; \frac{L}{2K}\, 10^{{\rm SNR}/10}
\;=\; G_{\rm eff}\, \gamma.
\end{equation}
\textbf{Theorem 1} (Effective processing gain). \emph{In the $K$-stream disjoint-support
sparse-code DSSS system with unit-norm columns, AWGN measured-power calibration, and a
single-pilot zero-forcing receiver, the effective processing gain at the data-decision stage is
$G_{\rm eff}=L/(2K)$.} For the parameters of this paper ($L\!=\!360$, $K\!=\!12$,
$G_{\rm eff}\!=\!15$) this is a $\sim\!14$~dB reduction relative to the classical $L$ prediction,
relaxing toward $L/K$ as $\gamma\!\to\!\infty$. The full composite-power derivation, the pilot-ZF
perturbation analysis, and the proof of Theorem~1 are given in the supplementary material.
\fi

\subsection{Closed-Form BER Expressions}\label{sec:an_ber}

\subsubsection{AWGN}\label{sec:an_awgn}
\ifextended
Because the ZF stage divides the despread data by the \emph{random} pilot estimate
$\hat h\!\sim\!\mathcal{CN}(1,\sigma^2)$ of~\eqref{eq:hat_h_decomp}, the post-correction
decision statistic $z=(a_{k,n}+\tilde n_{k,n})/\hat h$ is only \emph{conditionally}
Gaussian given $\hat h$. The AWGN bit-error probability is therefore properly written as an
expectation over the estimate,
\begin{equation}\label{eq:pb_awgn_exact}
P_b^{\rm AWGN}(\gamma)\;=\;\E_{\hat h}\!\big[\,P_b(\gamma\mid\hat h)\,\big],
\end{equation}
rather than by substituting a single effective SNR into the QPSK $\Q$-function. For
Gray-coded QPSK the quadrant decision on $z$ is invariant to the estimate magnitude
$|\hat h|$ and depends only on the pilot phase error $\phi\triangleq\arg\hat h$; conditioning
on $\phi$ and using $1/\sigma^2=(L/K)\gamma$ gives the conditional bit-error probability
\begin{equation}\label{eq:pb_awgn_cond}
\begin{split}
P_b(\gamma\mid\phi)=\;&\tfrac12\Q\!\Big((\cos\phi+\sin\phi)\sqrt{\tfrac{L}{K}\gamma}\Big)\\
&+\tfrac12\Q\!\Big((\cos\phi-\sin\phi)\sqrt{\tfrac{L}{K}\gamma}\Big),
\end{split}
\end{equation}
which is averaged over $\phi=\arg(1+\tilde n_p/p_0)$ to
evaluate~\eqref{eq:pb_awgn_exact}. This makes the role of the noisy channel estimate explicit:
in AWGN ($h=1$) the pilot penalty is a phase-error effect, and it vanishes as
$\sigma^2\!\to\!0$ (i.e.\ $\phi\!\to\!0$), where~\eqref{eq:pb_awgn_cond} recovers the
perfect-CSI value $\Q(\sqrt{(L/K)\gamma})$.

In the high-pilot-SNR regime ($\sigma^2\!\ll\!1$), a first-order expansion of the noisy-ZF
division (equivalently, absorbing the phase-error averaging into the effective $3$~dB noise
penalty of Section~\ref{sec:an_pilot}) collapses~\eqref{eq:pb_awgn_exact} to the compact
approximation
\begin{equation}\label{eq:pb_awgn}
P_b^{\rm AWGN}(\gamma)\;\approx\;\Q\!\left(\sqrt{G_{\rm eff}\,\gamma}\right)
\;=\;\Q\!\left(\sqrt{\tfrac{L\,\gamma}{2K}}\right),
\end{equation}
which is the baseline plotted in Fig.~\ref{fig:ana_awgn}. We stress
that~\eqref{eq:pb_awgn} is \emph{not} a universal closed form for sparse spreading: it is
the small-estimation-error limit of~\eqref{eq:pb_awgn_exact}, valid only under the
measured-power SNR convention of Section~\ref{sec:an_snrdef} and the single-pilot ZF receiver
of Section~\ref{sec:an_pilot}. Under these conditions it matches simulation within $1$~dB
across $\mathrm{SNR}\in[-20, -4]$~dB; the $\approx\!30\%$ lower simulated BER at
$\mathrm{SNR}\!\geq\!-2$~dB reflects the higher-order (phase-averaging) terms dropped in
passing from~\eqref{eq:pb_awgn_exact} to~\eqref{eq:pb_awgn}, consistent with the
$L/(2K)\!\to\!L/K$ transition noted after~\eqref{eq:gamma_eff_full}.
\else
Because the ZF stage divides by the noisy pilot estimate, the AWGN bit-error probability is
properly an expectation over the estimate; in the high-pilot-SNR limit this collapses, via the
effective $3$~dB penalty of Section~\ref{sec:an_snrdef}, to the compact baseline
\begin{equation}\label{eq:pb_awgn}
P_b^{\rm AWGN}(\gamma)\;\approx\;\Q\!\left(\sqrt{G_{\rm eff}\,\gamma}\right)
\;=\;\Q\!\left(\sqrt{\tfrac{L\,\gamma}{2K}}\right),
\end{equation}
plotted in Fig.~\ref{fig:ana_awgn}. This is the small-estimation-error limit (not a universal
closed form), valid under the measured-power SNR convention and the single-pilot ZF receiver; the
exact conditional-phase expectation is given in the supplementary material.
\fi

\subsubsection{Flat Rayleigh Fading with AWGN}\label{sec:an_rayleigh}
\ifextended
Under flat fading~\eqref{eq:ray_channel} the pilot no longer estimates a known unit channel:
from~\eqref{eq:hest} the estimate is $\hat h = h + \tilde n_p/p_0$ with
$\tilde n_p\!\sim\!\mathcal{CN}(0,\sigma^2)$, so the true channel $h$ and its estimate
$\hat h$ are jointly complex Gaussian with correlation
\begin{equation}\label{eq:rho_est}
\rho \;\triangleq\; \frac{\E[h\,\hat h^*]}{\sqrt{\E|h|^2\,\E|\hat h|^2}}
      \;=\; \frac{1}{\sqrt{1+\sigma^2}} .
\end{equation}
As in the AWGN case, the ZF detector divides by the \emph{random} estimate $\hat h$, so the
BER must be averaged over the joint law of $(h,\hat h)$, not over $|h|^2$ alone:
\begin{equation}\label{eq:pb_ray_exact}
P_b^{\rm Ray}(\gamma) \;=\; \E_{h,\hat h}\!\big[\,P_b(\gamma\mid h,\hat h)\,\big].
\end{equation}
Conditioned on the realized channel $h$ and estimate $\hat h$, the quadrant decision on
$z=(h\,a_{k,n}+\tilde n_{k,n})/\hat h$ depends on the true fading magnitude $|h|$ and the
\emph{estimation phase error} $\psi\triangleq\arg h-\arg\hat h$; using
$1/\sigma^2=(L/K)\gamma$,
\begin{equation}\label{eq:pb_ray_cond}
\begin{split}
P_b(\gamma\mid h,\hat h)=\;&\tfrac12\Q\!\Big(|h|(\cos\psi+\sin\psi)\sqrt{\tfrac{L}{K}\gamma}\Big)\\
&+\tfrac12\Q\!\Big(|h|(\cos\psi-\sin\psi)\sqrt{\tfrac{L}{K}\gamma}\Big),
\end{split}
\end{equation}
the direct fading analogue of the AWGN conditional law~\eqref{eq:pb_awgn_cond}, which it
recovers for $|h|\!=\!1$.

The closed-form baseline plotted in Fig.~\ref{fig:ana_awgn} follows, exactly as in the AWGN
case, by replacing the conditional law~\eqref{eq:pb_ray_cond} with a single effective
per-realization SNR $\gamma_{\rm inst}=|h|^2\,G_{\rm eff}\,\gamma$ (i.e., folding the
estimation phase penalty into the $3$~dB factor of $G_{\rm eff}$) and averaging
$\Q(\sqrt{|h|^2 G_{\rm eff}\gamma})$ over $|h|^2\!\sim\!\mathrm{Exp}(1)$
\cite[Eq.~(3.7)]{tse2005}:
\begin{equation}\label{eq:pb_rayleigh}
P_b^{\rm Ray}(\gamma) \;\approx\; \frac{1}{2}\!\left(1 - \sqrt{\frac{\mu}{1+\mu}}\right),
\quad \mu = \frac{G_{\rm eff}\,\gamma}{2}.
\end{equation}
Like~\eqref{eq:pb_awgn}, this is an approximation of the exact
expectation~\eqref{eq:pb_ray_exact}, not a general closed form: it collapses the estimation
phase error of~\eqref{eq:pb_ray_cond} into the fixed penalty $G_{\rm eff}$. The
error-free-estimate limit ($\rho\!\to\!1$, $\psi\!\to\!0$) of~\eqref{eq:pb_ray_cond} instead
removes the $3$~dB penalty and gives $\mu=(L/2K)\gamma$; retaining $G_{\rm eff}=L/(2K)$
(hence $\mu=L\gamma/(4K)$) is the effective-penalty form the noisy single-pilot receiver is
observed to follow across the operating range. At high SNR ($\mu\!\gg\!1$),
\eqref{eq:pb_rayleigh} yields $P_b^{\rm Ray}\!\approx\!1/(4\mu)=K/(L\gamma)$, the classical
first-order Rayleigh diversity slope.

\textit{Deep-fade regime.} The approximation~\eqref{eq:pb_rayleigh} is weakest when the fade is
comparable to the noise, $|h|\!\lesssim\!\sigma$: the pilot estimate is then noise-dominated, the
ZF ratio $h/\hat h$ becomes heavy-tailed (a quotient of complex Gaussians, with no finite second
moment) rather than a fixed $3$~dB noise scaling, and the phase error drives
$P_b(\gamma\mid h,\hat h)\!\to\!\tfrac12$ for the affected symbols, an error-floor contribution
rather than an SNR shift. The $|h|^2$-only form~\eqref{eq:pb_rayleigh} omits these effects and
under-predicts BER in the deep-fade tail, where it should be replaced by numerical evaluation of
the exact expectation~\eqref{eq:pb_ray_exact}.

Eq.~\eqref{eq:pb_rayleigh} further requires the noise variance to be \emph{fixed} in
absolute terms (independent of $h$), as adopted in our Rayleigh simulation per
Section~\ref{sec:sm_rayleigh}. Under the alternative measured-power model, the effective
per-frame SNR is held constant by automatic gain control (AGC)-like re-calibration, eliminating
the fading penalty.
\else
Under flat fading~\eqref{eq:ray_channel} the pilot estimate $\hat h=h+\tilde n_p/p_0$ is itself
noisy, so the BER must be averaged over the joint law of $(h,\hat h)$. Collapsing the estimation
phase error into the fixed $G_{\rm eff}$ penalty and averaging $\Q(\sqrt{|h|^2 G_{\rm eff}\gamma})$
over $|h|^2\!\sim\!\mathrm{Exp}(1)$~\cite[Eq.~(3.7)]{tse2005} gives the closed-form baseline
\begin{equation}\label{eq:pb_rayleigh}
P_b^{\rm Ray}(\gamma) \;\approx\; \frac{1}{2}\!\left(1 - \sqrt{\frac{\mu}{1+\mu}}\right),
\quad \mu = \frac{G_{\rm eff}\,\gamma}{2},
\end{equation}
which at high SNR yields the first-order Rayleigh diversity slope $K/(L\gamma)$. The exact
conditional-phase expectation, the deep-fade regime where~\eqref{eq:pb_rayleigh} under-predicts
BER, and the fixed-noise requirement are detailed in the supplementary material.
\fi

\subsubsection{Periodic Pulse Jamming on AWGN: Sparse $(M\!+\!1)$-State Model}\label{sec:an_pulse}

For sparse codes with $M$ non-zero entries per column, the classical two-state Bernoulli
pulse-jammer model~\cite{poisel2011}, which assumes the receiver-integrated jamming
power averages over many chips per symbol, substantially underestimates BER because only the
$M$ non-zero chip positions per column carry that stream's signal energy: a pulse hit on a
zero-position has no effect, while a pulse hit on a non-zero position contributes
disproportionately to the despread output. We refine it into an
$(M\!+\!1)$-state model that enumerates the number of non-zero chip positions hit by the
pulse during a given symbol.

\ifextended
\textit{Derivation.}
Assume, as an approximation to the periodic pulse jammer~\eqref{eq:pulse_jammer}, that each
non-zero chip position is hit independently with probability $\rho$ (its duty cycle). For
column $\bc_k$ with $M$ non-zero positions, the random number of hits $i\in\{0,\dots,M\}$ then
follows a binomial distribution, which can be described by
\begin{equation}\label{eq:p_states_general}
P_i \;=\; \binom{M}{i}\,\rho^i\,(1-\rho)^{M-i}.
\end{equation}
This is consistent with~\eqref{eq:pulse_jammer} \emph{in the marginal}: the ON-state indicator
of~\eqref{eq:pulse_jammer} has duty cycle $\rho$, so each individual non-zero chip is struck
with marginal probability $\rho$, matching the per-chip hit rate of the jammer. The
\emph{independence} across the $M$ positions is the approximating step. Equation
\eqref{eq:pulse_jammer} is deterministic and periodic with a contiguous ON-interval of duration
$\rho/f_p$; hit events are therefore in general correlated, through the shared pulse phase and,
for closely spaced positions, through a single ON-burst spanning several consecutive chips.
Independence is recovered in the limit where the pulse phase is uniform relative to the code
grid and the non-zero positions are spaced wider than the ON-interval, so that no burst covers
two of them; the correlated regime and its effect on~\eqref{eq:pb_pulse} are quantified under
\emph{Model assumptions and scope} below.
Conditional on $i$ hits, the despread output gains an additional jamming-power contribution
$\sigma_J^2 = (i/M)\,J_{\rm on}$ where $J_{\rm on}=\mathrm{JSR}\,\sigma^2/\rho$ is the
on-state jammer power per chip. The $1/\rho$ factor compensates for the $\rho$-fraction
of time the jammer is on, giving average power $\mathrm{JSR}\,\sigma^2$. Combining with
the AWGN noise variance $\sigma^2$ and applying the effective processing gain $G_{\rm eff}$,
the state-conditioned post-ZF SNR is
\begin{equation}\label{eq:gammai}
\gamma_i \;=\; \frac{G_{\rm eff}\,\gamma}{1 + (i/M)\,\mathrm{JSR}/\rho},
\qquad i\in\{0,1,\dots,M\}.
\end{equation}

For QPSK with these state-conditional SNRs, the total bit-error probability is the
binomial-weighted average, which can be obtained as
\begin{equation}\label{eq:pb_pulse}
P_b^{\rm pulse}(\gamma,\mathrm{JSR}) \;=\; \sum_{i=0}^{M}\binom{M}{i}\rho^i(1-\rho)^{M-i}\,\Q\!\left(\sqrt{\gamma_i}\right).
\end{equation}

\textit{Special case $M=2$ (the recommended sparse design):} For three states
$i\!\in\!\{0,1,2\}$, \eqref{eq:pb_pulse} expands to
\begin{align}
P_b^{\rm pulse}|_{M=2} \;=\;& (1-\rho)^2\,\Q\!\left(\sqrt{G_{\rm eff}\,\gamma}\right) \nonumber\\
&\;+\; 2\rho(1-\rho)\,\Q\!\left(\sqrt{\frac{G_{\rm eff}\,\gamma}{1+\mathrm{JSR}/(2\rho)}}\right) \nonumber\\
&\;+\; \rho^2\,\Q\!\left(\sqrt{\frac{G_{\rm eff}\,\gamma}{1+\mathrm{JSR}/\rho}}\right).
\label{eq:pb_pulse_M2}
\end{align}
At low duty cycle ($\rho \ll 1$), $P_b^{\rm pulse}|_{M=2}\!\approx\! P_b^{\rm AWGN}$, recovering
the unjammed limit. As $\rho\!\to\!1$, $P_b^{\rm pulse}|_{M=2}\!\to\!\Q(\sqrt{G_{\rm eff}\gamma/(1+\mathrm{JSR})})$,
recovering the always-jammed AWGN+JSR limit. Between these extremes, the
binomial mixture captures the sparse-code-specific behavior that the classical
two-state model misses.

\textit{Model assumptions and scope.} The binomial law~\eqref{eq:p_states_general} treats the
$M$ hits as independent with duty-cycle probability $\rho$, which holds when the pulse on-time is
short relative to the spacing between a column's non-zero chips, or equivalently when the pulse
phase is uniform from symbol to symbol. It is therefore an \emph{ensemble} model for the
JSR-averaged BER of a spectrally flat code and does not resolve two regimes: (i) \emph{long
(correlated) pulses}, where one pulse strikes several consecutive non-zero chips, making the hits
bursty and sub-binomial so that the tails of~\eqref{eq:pb_pulse} broaden (a Gilbert--Elliott
block model would be needed, and the independent-hit form is its decorrelated limit); and (ii)
\emph{deterministic spectral alignment}, namely the Paley--QR failure (Section~\ref{sec:res_algebraic}),
a frequency-domain resonance in which the quadratic-residue spectral peaks coincide with the
$10$~kHz pulse harmonics, a mechanism invisible to the flat-spectrum model~\eqref{eq:pb_pulse}
and characterized separately via Fig.~\ref{fig:spectral_compare}.

\textit{Comparison to the classical Bernoulli model:}
The two-state classical Bernoulli model is the $M\!\to\!\infty$ limit of~\eqref{eq:pb_pulse},
where the law of large numbers averages the binomial mixture into a single weighted-noise floor.
In this case,
\begin{equation}\label{eq:pb_bernoulli_classical}
P_b^{\rm classical} \;=\; \rho\,\Q\!\left(\sqrt{\frac{G_{\rm eff}\gamma}{1+\mathrm{JSR}/\rho}}\right) + (1-\rho)\,\Q\!\left(\sqrt{G_{\rm eff}\gamma}\right).
\end{equation}
For $M\!=\!2$, the difference between~\eqref{eq:pb_pulse_M2} and~\eqref{eq:pb_bernoulli_classical}
is $O(\rho^2)$ per~\eqref{eq:pb_pulse_M2}'s third term (the quadratic-in-$\rho$ ``both hit''
state). The discrepancy is one of our
analytical contributions: \emph{the sparse pulse model corrects the classical Bernoulli
expression in the $M\!\to\!1$ limit, where binomial finite-$M$ effects matter}.
\else
Assuming each non-zero chip is hit independently with the duty-cycle probability $\rho$, the
number of hits on a column's $M$ positions is binomial, and the QPSK bit-error probability is the
binomial-weighted average over the $(M\!+\!1)$ hit states:
\begin{equation}\label{eq:pb_pulse}
\begin{aligned}
P_b^{\rm pulse}(\gamma,\mathrm{JSR}) &\;=\; \sum_{i=0}^{M}\binom{M}{i}\rho^i(1-\rho)^{M-i}\,\Q\!\left(\sqrt{\gamma_i}\right),\\
\text{with}\;\; \gamma_i &\;=\;\frac{G_{\rm eff}\,\gamma}{1+(i/M)\,\mathrm{JSR}/\rho}.
\end{aligned}
\end{equation}
For the recommended $M\!=\!2$ design this reduces to the three-state form
\begin{align}
P_b^{\rm pulse}|_{M=2} \;=\;& (1-\rho)^2\,\Q\!\left(\sqrt{G_{\rm eff}\,\gamma}\right) \nonumber\\
&+\; 2\rho(1-\rho)\,\Q\!\left(\sqrt{\tfrac{G_{\rm eff}\,\gamma}{1+\mathrm{JSR}/(2\rho)}}\right) \nonumber\\
&+\; \rho^2\,\Q\!\left(\sqrt{\tfrac{G_{\rm eff}\,\gamma}{1+\mathrm{JSR}/\rho}}\right),
\label{eq:pb_pulse_M2}
\end{align}
whose classical two-state Bernoulli limit ($M\!\to\!\infty$) is
\begin{equation}\label{eq:pb_bernoulli_classical}
P_b^{\rm classical} \;=\; \rho\,\Q\!\left(\sqrt{\tfrac{G_{\rm eff}\gamma}{1+\mathrm{JSR}/\rho}}\right) + (1-\rho)\,\Q\!\left(\sqrt{G_{\rm eff}\gamma}\right).
\end{equation}
The sparse model corrects this classical expression by $O(\rho^2)$ in the finite-$M$ regime. The
full derivation, the independence assumption, and the deterministic spectral-alignment caveat
(the Paley--QR resonance) are given in the supplementary material.
\fi

\subsection{Cross-Talk in the Overlap Regime}\label{sec:an_crosstalk}

With $K$
columns each holding $M$ non-zeros, the total occupancy $KM$ cannot exceed $L$
if columns are to remain disjoint. The bound $KM\!\leq\!L$ is tight at
$M\!=\!L/K$ (equal to $30$ for our NB-IoT parameters $L\!=\!360$, $K\!=\!12$),
beyond which disjoint allocation is combinatorially infeasible. For random
allocation beyond this threshold, a standard collision-probability approximation gives the
per-pair overlap probability
\begin{equation}\label{eq:p_overlap}
P_{\rm ovr}(M) \;\approx\; 1\!-\!\left(1\!-\!\tfrac{M}{L}\right)^M,
\end{equation}
which transitions from $\ll\!1$ to $\sim\!1$ as $M$ crosses
$\sqrt{L}\!\approx\!19$ for our parameters (well before $L/K\!=\!30$)
and saturates at $1$ for $M\!\geq\!L/K$. The implication is that the
$M\!\approx\!L/K$ transition is not a coincidence of the simulation
parameters but a structural threshold dictated by the
$(L,K,M)$ combinatorics.

\ifextended
For the overlap regime where disjoint supports are infeasible, the despread
output~\eqref{eq:despread} carries an inter-column cross-talk term whose magnitude is
governed by the off-diagonal entries of the Gram matrix $\bC^H\bC$. For columns drawn with
i.i.d.\ unit-norm complex-Gaussian values on independently-chosen $M$-subsets of
$\{1,\dots,L\}$, the expected total squared correlation obeys the Welch-type
estimate~\cite{welch1974}
\begin{equation}\label{eq:tsc_overlap}
\E\!\left[\sum_{i\neq j}|\bc_i^H\bc_j|^2\right] \;\approx\; \frac{K(K-1)\,M^2}{L^2}.
\end{equation}
Eq.~\eqref{eq:tsc_overlap} is a leading-order estimate for moderate overlap and is
\emph{monotonically increasing} in $M$; it captures the onset and initial growth of
inter-column cross-talk once disjoint supports become infeasible ($M\!>\!L/K$), but it does
not by itself predict a peak or a subsequent decay. The non-monotone BER measured in
Section~\ref{sec:res_density} (a sparse-regime minimum ($M\!=\!2$--$5$), a bad-zone peak at
$M\!\approx\!60$--$120$, and a partial recovery for $M\!\geq\!180$) arises instead from the
\emph{interplay} between this rising cross-talk and the per-column averaging that suppresses
it. Just past the disjoint threshold the overlaps are few and concentrated, so the cross-talk
adds coherently and inflates BER; at high density each column integrates many non-zero chips,
averaging the same cross-talk toward a smaller noise-like residual while the per-column signal
budget grows. Thus~\eqref{eq:tsc_overlap} and the overlap probability~\eqref{eq:p_overlap} fix
the \emph{location} of the bad-zone onset at $M\!=\!L/K$, whereas the recovery at large $M$ is
an averaging effect beyond this leading-order estimate; a quantitative peak-and-recovery model
is left to future analytical refinement.
\else
For overlapping supports the despread output~\eqref{eq:despread} carries an inter-column
cross-talk term. For columns with i.i.d.\ unit-norm values on random $M$-subsets, the expected
total squared correlation obeys the Welch-type estimate~\cite{welch1974}
\begin{equation}\label{eq:tsc_overlap}
\E\!\left[\sum_{i\neq j}|\bc_i^H\bc_j|^2\right] \;\approx\; \frac{K(K-1)\,M^2}{L^2},
\end{equation}
which is monotonically increasing in $M$ and fixes the \emph{onset} of cross-talk at
$M\!=\!L/K$. The non-monotone bad-zone BER (peak at $M\!\approx\!60$--$120$, partial recovery for
$M\!\geq\!180$; Section~\ref{sec:res_density}) arises from the interplay between this rising
cross-talk and the per-column averaging that suppresses it; the detailed derivation is given in
the supplementary material.
\fi

\section{Numerical Results}\label{sec:results}

This section presents simulations that (i) validate the analytical framework, and (ii)
characterize the design space. Subsection~\ref{sec:res_setup} describes the simulation
configuration. Subsection~\ref{sec:res_analytical} reports the analytical-vs-simulation overlay.
Subsections~\ref{sec:res_density}--\ref{sec:res_doppler} present the empirical findings on density,
structure, algebraic constructions, ML approaches, and Doppler. Subsection~\ref{sec:res_final}
summarizes the recommended designs across all four threat regimes.

\subsection{Simulation Configuration}\label{sec:res_setup}

The simulator is implemented in MATLAB using the LTE Toolbox, with NB-IoT parameters
following~\cite{cheng2022, etsi36101, etsi36104}. The simulation parameters are summarized in
\ifextended Table~\ref{tab:sim_params}\else the supplementary material\fi. Each BER point is averaged over a deterministic seed-set of frames and
reported with $95\%$ Wilson-score confidence intervals where applicable; pilot-aided ZF
equalization uses a single pilot symbol at $(k,n)=(1,1)$, with $\hat h$ computed
from~\eqref{eq:hest} and applied via~\eqref{eq:zf}, and AJM is computed by linear interpolation of
$\log_{10}(\mathrm{BER})$.

\ifextended
\begin{table}[!t]
\centering
\caption{Simulation parameters.}
\label{tab:sim_params}
\small
\begin{tabular}{@{}l@{\hspace{6pt}}p{0.56\columnwidth}@{}}
\toprule
Parameter & Value \\
\midrule
Spreading length $L$ & $360$ chips per data symbol \\
Subcarriers $K$ & $12$ \\
Symbols per subframe $N_s$ & $14$ \\
Modulation & QPSK \\
Frames per BER point & $20$--$1000$ \\
Confidence intervals & $95\%$ Wilson score \\
Fading channel & 3GPP ETU, $f_d\!\in\!\{5,70,200,500\}$~Hz \\
Pilot / equalization & single pilot $p_0\!=\!1$ at $(1,1)$; ZF \\
Pulse jammer $(f_p,\rho)$ & $(10\,\text{kHz},2\%)$, $(20\,\text{kHz},2\%)$, $(50\,\text{kHz},1\%)$ \\
Multi-tone jammer $K_t$ & $\{1,2,3,4\}$ tones \\
AJM threshold $\mathrm{BER}_{\rm th}$ & $10^{-2}$ \\
\bottomrule
\end{tabular}
\end{table}
\fi

\textit{Equal-power assumption and scope.} All $K\!=\!12$ streams are modeled with equal
received power (unit-energy symbols and unit-norm columns, giving the composite power
$P_x\!=\!K/L$ of Section~\ref{sec:an_snrdef}), so the disjoint-support designs carry zero
inter-stream interference and the overlap-regime cross-talk is treated as an aggregate noise
contribution.
\ifextended
This is appropriate for the
synchronous, power-controlled NB-IoT downlink considered here, but it does not capture the
near--far regime of dense NOMA/SCMA deployments: with unequal user powers $\{P_j\}$ the
interference seen at column $k$ is the power-weighted $\sum_{j\neq k}P_j\,|\bc_j^H\bc_k|^2$
rather than a uniform aggregate, so the cross-correlation impact (and any near--far
disparity) becomes strongly dependent on the user power profile. Extending the density-first
analysis to unequal-power and asynchronous NOMA/SCMA profiles is left to future work.
\else
This suits the synchronous, power-controlled NB-IoT downlink considered here; the near--far
unequal-power regime of dense NOMA/SCMA deployments, where the interference at column $k$
becomes the power-weighted $\sum_{j\neq k}P_j\,|\bc_j^H\bc_k|^2$, is left to future work.
\fi

\subsection{Analytical Validation}\label{sec:res_analytical}

We first validate the closed-form analytical expressions of Section~\ref{sec:analytical} on
the recommended uniform-$M\!=\!2$ Free code (the design that the rest of the section
identifies as the optimal sparse choice). The $1000$-frame simulation comprises
$334{,}000$ payload bits per BER point.

\ifextended
Fig.~\ref{fig:ana_validation} overlays the closed-form curves on simulation for the AWGN~(a),
flat-Rayleigh~(b), and periodic-pulse~(c) cases. The
analytical curves (solid black) are computed from~\eqref{eq:pb_awgn},~\eqref{eq:pb_rayleigh},
and~\eqref{eq:pb_pulse}; simulation points (blue dashed line, filled circles) are shown
with $95\%$ Wilson-score CI error bars. Across the meaningful operating range
($\mathrm{BER}\in[10^{-4}, 10^{-1}]$), simulation tracks analytical within $1$--$2$~dB,
validating the $G_{\rm eff} = L/(2K)$ correction. The minor discrepancies reflect (i) the SNR-dependent ZF penalty, (ii) deep-fade
undersampling at high SNR, and (iii) periodic-pulse position correlation, all consistent
with the analytical model's stated approximations.

For the periodic-pulse case, the difference between the sparse $(M\!+\!1)$-state
model~\eqref{eq:pb_pulse} and the classical two-state Bernoulli
expression~\eqref{eq:pb_bernoulli_classical} is $O(\rho^2)$: a small effect at $\rho\!=\!2\%$ but
a large one on the analytical-vs-simulation agreement at high JSR. The Rayleigh validation, in
turn, requires a \emph{fixed}-noise simulation mode: a simulator that applies
\texttt{awgn(`measured')} on the post-fading waveform holds the per-frame SNR constant and
produces results indistinguishable from AWGN, masking the true Rayleigh impact.

\begin{figure*}[!t]
\centering
\subcaptionbox{AWGN~\eqref{eq:pb_awgn}\label{fig:ana_awgn}}{%
   \includegraphics[width=0.32\textwidth]{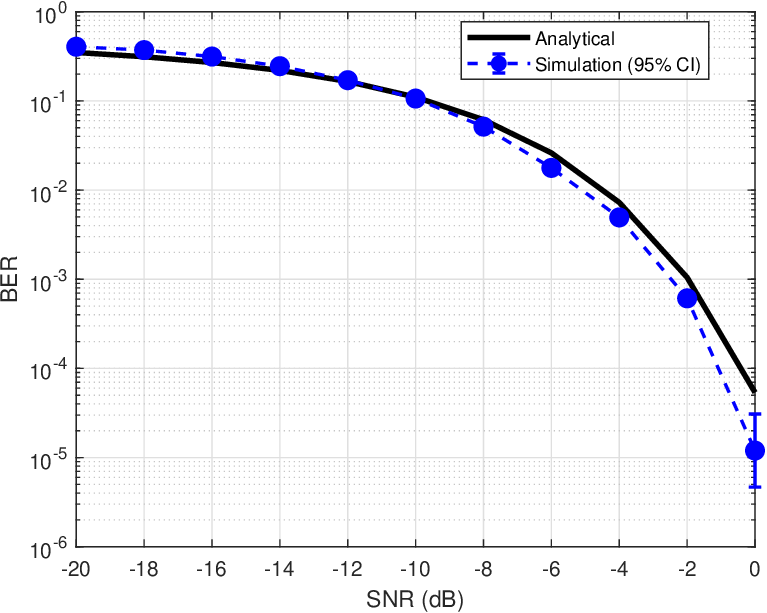}}\hfill
\subcaptionbox{Flat Rayleigh~\eqref{eq:pb_rayleigh}\label{fig:ana_rayleigh}}{%
   \includegraphics[width=0.32\textwidth]{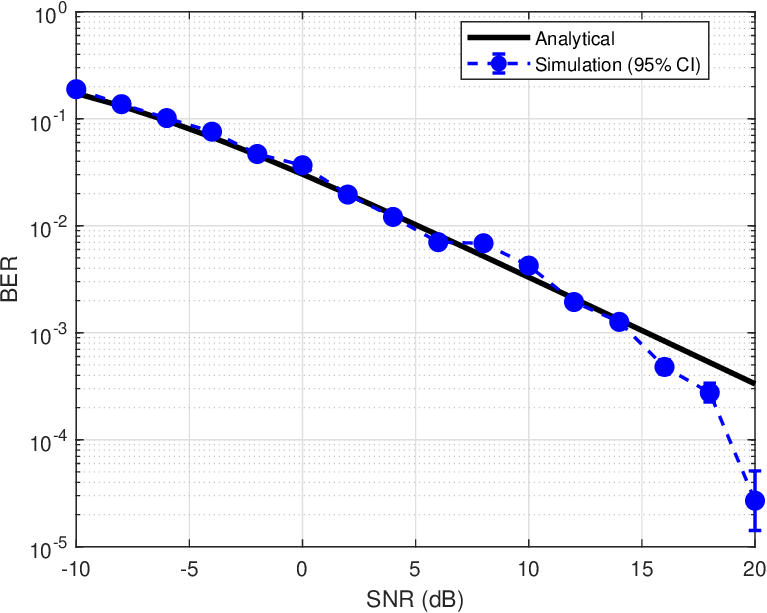}}\hfill
\subcaptionbox{Periodic pulse, $3$-state~\eqref{eq:pb_pulse}\label{fig:ana_pulse}}{%
   \includegraphics[width=0.32\textwidth]{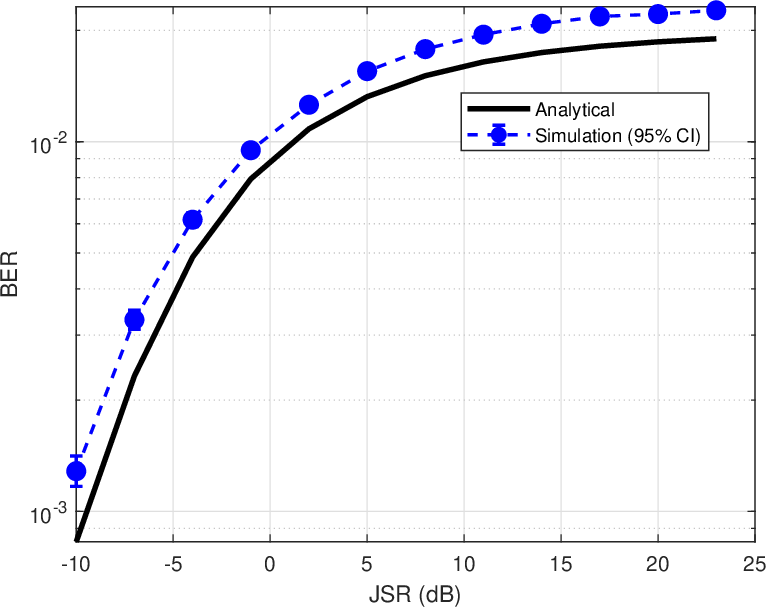}}
\caption{Closed-form analytical BER vs.\ simulation for the uniform $M\!=\!2$ Free design
($95\%$ Wilson-score CI): (a) AWGN; (b) flat Rayleigh in fixed-noise mode; (c) periodic
pulse ($f_p\!=\!10$~kHz, $\rho\!=\!2\%$). Simulation tracks analysis within $1$--$2$~dB,
validating the $G_{\rm eff}=L/(2K)$ correction.}
\label{fig:ana_validation}
\end{figure*}
\else
Fig.~\ref{fig:ana_awgn} overlays the closed-form AWGN curve~\eqref{eq:pb_awgn} on simulation
(blue dashed, filled circles, $95\%$ Wilson-score CI error bars). Across the meaningful
operating range ($\mathrm{BER}\in[10^{-4}, 10^{-1}]$) simulation tracks analysis within
$1$--$2$~dB, validating the $G_{\rm eff}=L/(2K)$ correction. The flat-Rayleigh and
periodic-pulse validations, together with a discussion of the residual discrepancies (the
SNR-dependent ZF penalty, deep-fade undersampling, and the $O(\rho^2)$ gap between the sparse
$(M\!+\!1)$-state and classical Bernoulli pulse models), are given in the supplementary material.

\begin{figure}[!t]
\centering
\includegraphics[width=\columnwidth]{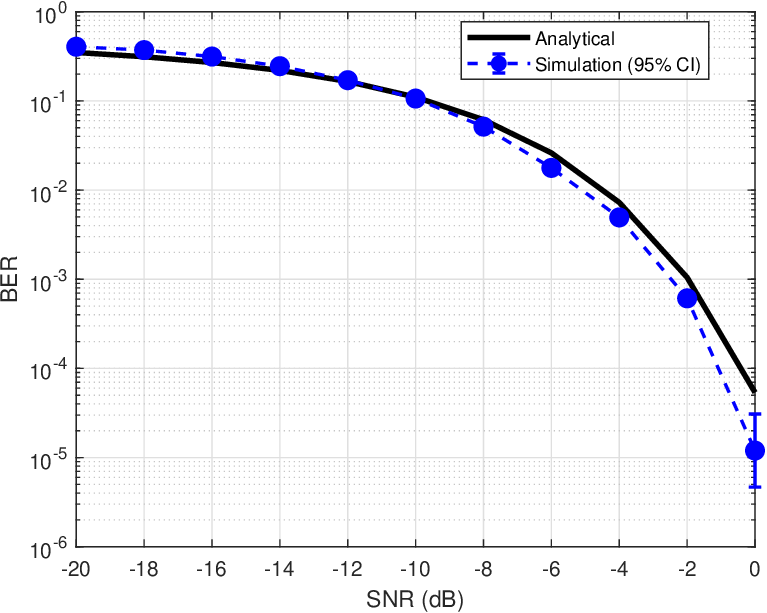}
\caption{Closed-form analytical BER vs.\ simulation in AWGN for the uniform $M\!=\!2$ Free
design ($95\%$ Wilson-score CI), validating the $G_{\rm eff}=L/(2K)$ correction. The
flat-Rayleigh and periodic-pulse validations are in the supplementary material.}
\label{fig:ana_awgn}
\end{figure}
\fi

\subsection{Density-Axis Sweep and the Bad-Zone Regime}\label{sec:res_density}

We sweep the per-column density $M$ from $2$ (extreme sparse) to $360$ (fully dense), with
$M\in\{2,5,10,20,30,60,120,180,240,300,360\}$. For each $M\!\leq\!L/K\!=\!30$ we use the
disjoint Free-Uniform construction; for $M\!>\!30$ we use the Free-Overlap construction
that allows column-support overlap.

\begin{figure}[!t]
\centering
\includegraphics[width=\columnwidth]{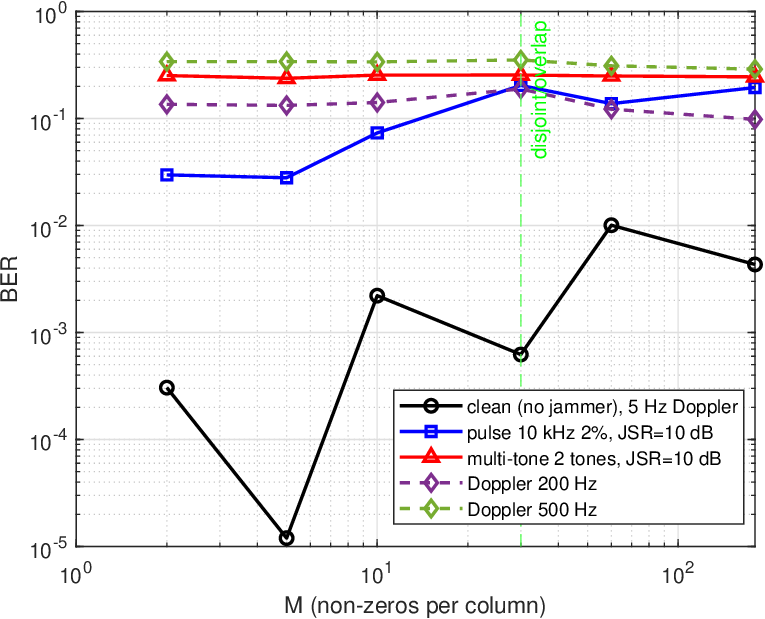}
\caption{BER vs.\ column density $M$ under clean, pulse, and multi-tone jamming. The dashed line
marks the disjoint/overlap boundary $M\!=\!L/K\!=\!30$; the bad zone appears at
$M\!\approx\!60$--$120$.}
\label{fig:density_sweep}
\end{figure}

Fig.~\ref{fig:density_sweep} presents the central empirical result of this paper. Three
operating curves are overlaid: clean (no jammer at SNR$\!=\!5$~dB), pulse jamming
($10$~kHz, $2\%$, JSR$\!=\!10$~dB), and multi-tone jamming ($2$ tones, JSR$\!=\!10$~dB).
The vertical dashed line at $M\!=\!L/K\!=\!30$ marks the disjoint/overlap boundary.

The figure reveals three distinct regimes:

\textit{Sparse regime} ($M\!=\!2$--$5$): pulse BER reaches $0.024$ at $M\!=\!5$, the
empirical minimum. Multi-tone BER is also at its minimum at $M\!=\!2$ ($0.20$). Disjoint
column supports eliminate inter-column cross-talk, leaving only intra-column noise to
defeat the receiver.

\textit{Bad zone} ($M\!\approx\!60$--$120$): immediately past the disjoint/overlap
boundary, the pulse BER jumps to $0.30$ ($\!\approx\!10\!\times$ the sparse-regime
optimum). Cross-column overlap introduces inter-column cross-talk per~\eqref{eq:tsc_overlap},
but the column averaging required to suppress it is not yet available because density is
still moderate.

\textit{Dense regime} ($M\!\geq\!180$): pulse BER recovers to $0.15$--$0.18$ as cross-talk
gets averaged across the many non-zero chips per column. Doppler performance is
dominantly determined by per-column averaging budget and is best at full density.

The LDSC paper~\cite{yang2024} operating point of $M\!=\!30$ via Alamouti pairing
(yielding 60 non-zeros per column) sits exactly in the upper edge of the bad zone. This
finding explains the relatively poor pulse-jamming performance reported there
relative to a properly sparse design.

\subsection{Free Family vs. Alamouti-Paired LDSC}\label{sec:res_free}

We compare the Free family against the density-matched Alamouti-paired LDSC
of~\cite{yang2024}, both at $30$ non-zeros per column. Averaged over seven jamming scenarios
(three pulse and four multi-tone configurations) at JSR$\!=\!10$~dB, the Free-Uniform
construction achieves a mean BER of $0.18$ versus $0.23$ for the Alamouti-paired construction,
a $\sim\!20\%$ reduction. The improvement is largest at pulse jamming, where the Free
construction's disjoint column supports eliminate the inter-column cross-talk that plagues the
paired construction.

\subsection{Structural Placement Comparison}\label{sec:res_structural}

\ifextended
Within the sparse Free regime ($M\!=\!2$, $3$, $5$), four structural placement variants are
compared: random, uniform, golden-ratio, and chirp. Across $1000$-frame BER measurements at
JSR$\!=\!10$~dB, the structural variants differ by $1$--$2$~dB in pulse jamming AJM. The
uniform-spacing variant emerges as a marginally best deterministic default, although the
differences are within MC variance for several of the comparisons. We adopt
uniform $M\!=\!2$ as the recommended Free-family configuration for the remainder of the
paper.
\else
Among the sparse Free placements (random, uniform, golden-ratio, chirp), uniform spacing is the
marginally best deterministic default (the differences are within Monte Carlo variance); we adopt
uniform $M\!=\!2$ as the recommended Free-family configuration. Details are in the supplementary
material.
\fi

\subsection{Algebraic Constructions: Lempel--Costas versus Paley-QR}\label{sec:res_algebraic}

\begin{figure*}[!t]
\centering
\subcaptionbox{Pulse jamming ($10$~kHz, $2\%$ duty)\label{fig:alg_pulse}}{%
   \includegraphics[width=0.33\textwidth]{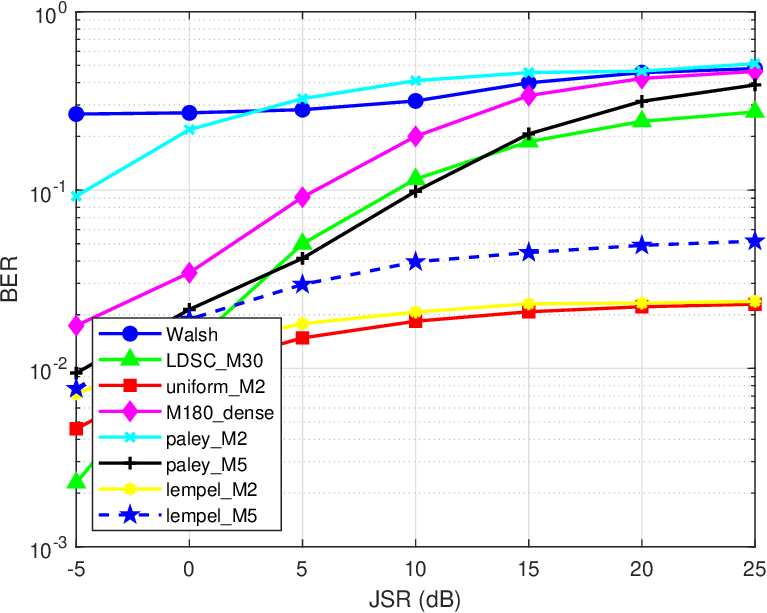}}\hfill
\subcaptionbox{Multi-tone jamming ($2$ tones)\label{fig:alg_mt}}{%
   \includegraphics[width=0.33\textwidth]{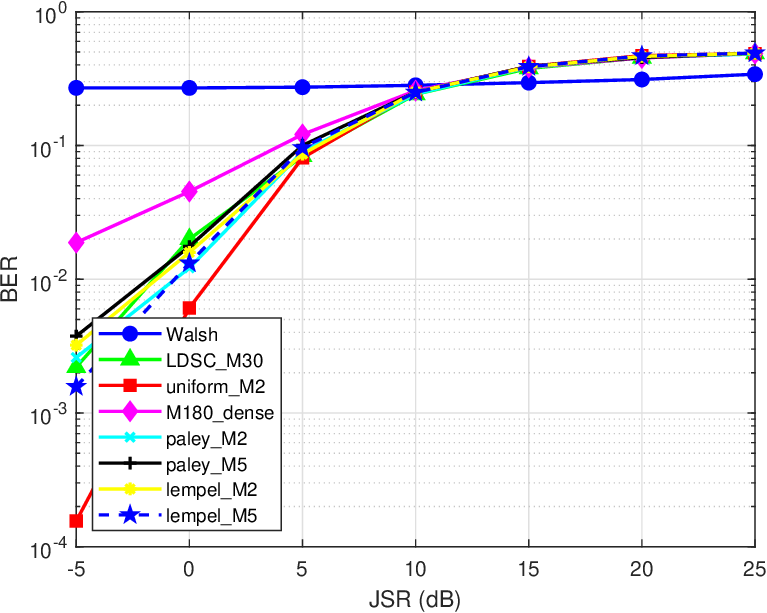}}\hfill
\subcaptionbox{ETU + $200$~Hz Doppler\label{fig:alg_dop}}{%
   \includegraphics[width=0.33\textwidth]{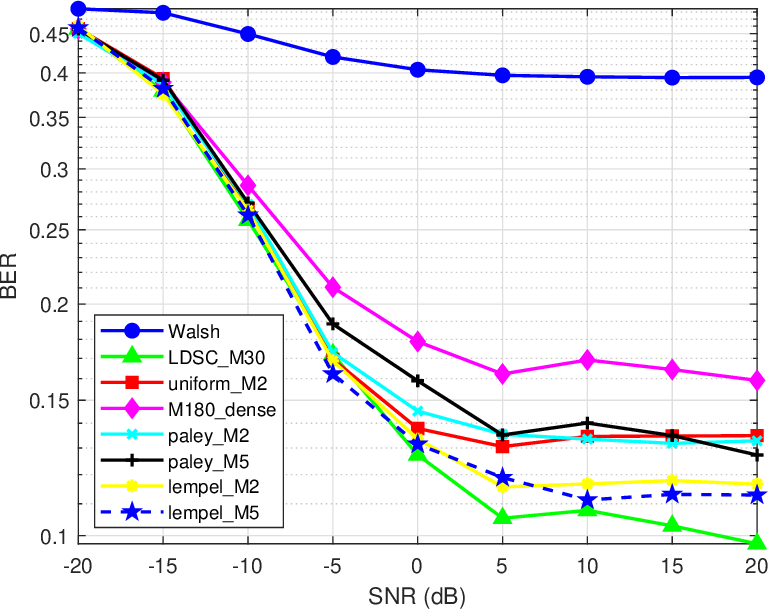}}
\caption{Algebraic constructions vs.\ baselines under (a) pulse, (b) multi-tone, and (c) ETU
$200$~Hz Doppler jamming.}
\label{fig:algebraic_comparison}
\end{figure*}

Algebraic constructions provide deterministic, reproducible alternatives to the random
free family. We evaluate Paley-QR and Lempel-Costas at $M\!=\!2$ and $M\!=\!5$, alongside
Walsh and the original LDSC-$M30$ as baselines.
Fig.~\ref{fig:algebraic_comparison} shows the BER vs.\ JSR for these eight codes
across pulse jamming (a), multi-tone jamming (b), and ETU + 200~Hz Doppler (c).

The headline finding is that \emph{Paley-QR at $M\!=\!2$ degrades to the random-decision floor at all JSR
levels} (BER $\!\approx\!0.48$), while \emph{Lempel-Costas at $M\!=\!2$ matches the
empirical optimum} (BER $\!=\!0.020$ at JSR$\!=\!10$~dB, vs.\ uniform $M\!=\!2$'s $0.018$).
\ifextended
The Paley-QR failure is attributable to the spectral structure of the QR set. Because a sparse
$M\!=\!2$ column is a two-tap sequence, its chip-rate spectrum is a raised cosine whose
peak-to-average ratio is fixed (near $4$~dB) for \emph{every} $M\!=\!2$ code; the distinguishing
feature is therefore not the peak height but the peak \emph{spacing}, set by the gap $\Delta$
between the two non-zero positions, with peaks recurring every $f_s/\Delta$ in frequency
($f_s$ the chip rate). The quadratic-residue
positions mod $359$ produce closely spaced spectral peaks that overlap the harmonic comb of the
$10$~kHz periodic pulse, coupling the jammer energy into the despread output, a resonance.
Lempel--Costas, by its log-domain construction, produces a coarser, offset spectrum whose peaks
largely fall between the jammer harmonics, avoiding the resonance.
Figure~\ref{fig:spectral_compare} shows this directly: the Paley-QR peaks recur several times more
finely than the Lempel--Costas and Free-uniform peaks and coincide with the $10$~kHz harmonics
(marked), explaining the catastrophic BER inflation in Fig.~\ref{fig:algebraic_comparison}(a).

\begin{figure}[!t]
\centering
\includegraphics[width=\columnwidth]{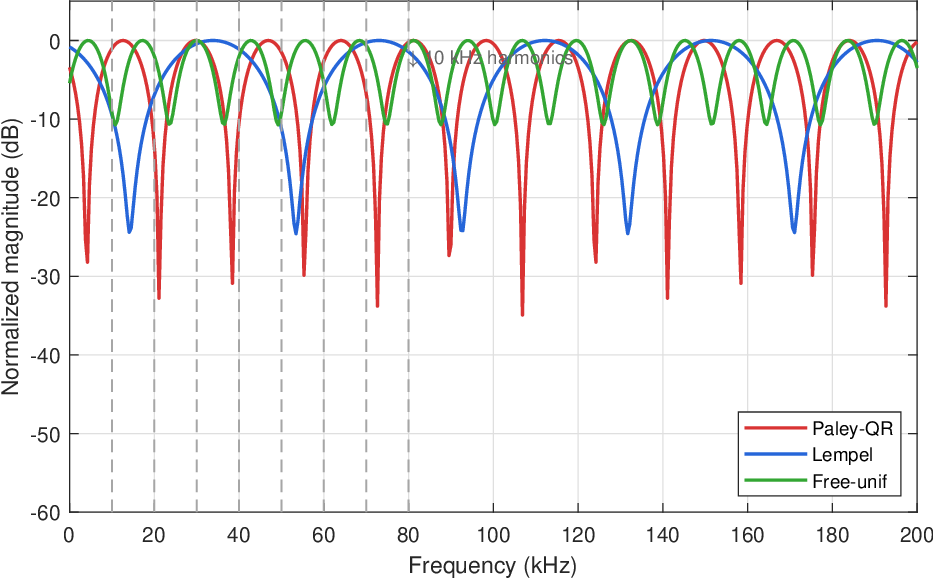}
\caption{Chip-rate FFT magnitude of representative $M\!=\!2$ columns. All two-tap columns share
the same peak-to-average ratio ($\approx\!4$~dB); the constructions differ in peak \emph{spacing}.
The Paley-QR peaks (red) recur finely and coincide with the $10$~kHz pulse harmonics (dashed),
whereas the Lempel--Costas (blue) and Free-uniform (green) peaks are coarser and fall between them.}
\label{fig:spectral_compare}
\end{figure}
\else
The Paley-QR failure is a deterministic spectral resonance: because an $M\!=\!2$ column is a
two-tap sequence with a fixed ($\approx\!4$~dB) peak-to-average ratio, the distinguishing feature
is the peak \emph{spacing}, and the quadratic-residue positions produce finely spaced peaks that
coincide with the $10$~kHz pulse harmonic comb, coupling the jammer into the despread output;
Lempel--Costas avoids this. The chip-rate spectra are shown in the supplementary material.
\fi

The Lempel-Costas $M\!=\!2$ result (Fig.~\ref{fig:algebraic_comparison}) demonstrates that a
closed-form, deterministic algebraic construction can achieve the same jamming-robustness
performance as the empirically-optimal Free design, with the practical advantage of full
reproducibility.

\subsection{Predictive and Generative Machine Learning}\label{sec:res_ml}

We evaluate both predictive and generative ML approaches per the methodology
of Subsection~\ref{sec:fw_ml}.

\subsubsection{Predictive ML}
\ifextended
Figure~\ref{fig:nn_predictor_fail} shows the predictor performance scatter: the trained
neural-network regressor (and an alternative random-forest regressor) achieve test
$R^2 < 0$ on held-out data, indicating that predictions are worse than predicting the
test mean. The cause is that the per-matrix wBER variance from
finite-sample Monte Carlo simulation overwhelms the signal at $N\!=\!210$ training samples.

\begin{figure}[!t]
\centering
\includegraphics[width=\columnwidth]{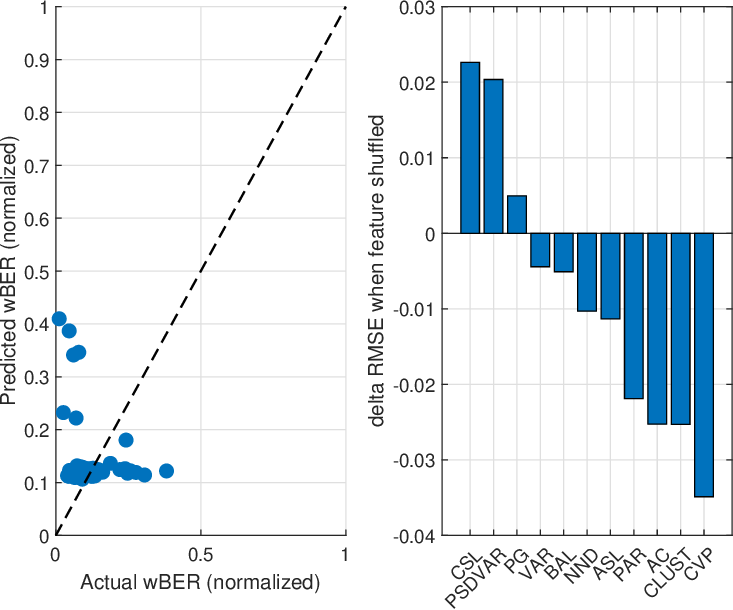}
\caption{Predictive-ML failure: predicted vs.\ actual wBER (left) and permutation feature
importance (right); test $R^2<0$ on the $210$-matrix set.}
\label{fig:nn_predictor_fail}
\end{figure}
\else
Both a trained neural-network regressor and a random-forest regressor achieve test $R^2<0$ on
held-out data (worse than predicting the mean): the per-matrix wBER variance from finite-sample
Monte Carlo overwhelms the signal at $N\!=\!210$ samples. The predictor scatter and
permutation-importance plots are in the supplementary material.
\fi

\subsubsection{Generative ML}

In contrast, the linear VAE generator produces sparse codes
that match or exceed the best deterministic algebraic construction across all three threat
regimes.
\ifextended
Fig.~\ref{fig:vae_comparison} compares VAE-best vs.\ Lempel-Costas $M\!=\!2$
vs.\ uniform $M\!=\!2$ in the multi-threat operating regime at $1000$ frames per point.

\begin{figure*}[!t]
\centering
\subcaptionbox{Pulse jamming ($10$~kHz, $2\%$ duty)\label{fig:vae_pulse}}{%
   \includegraphics[width=0.33\textwidth]{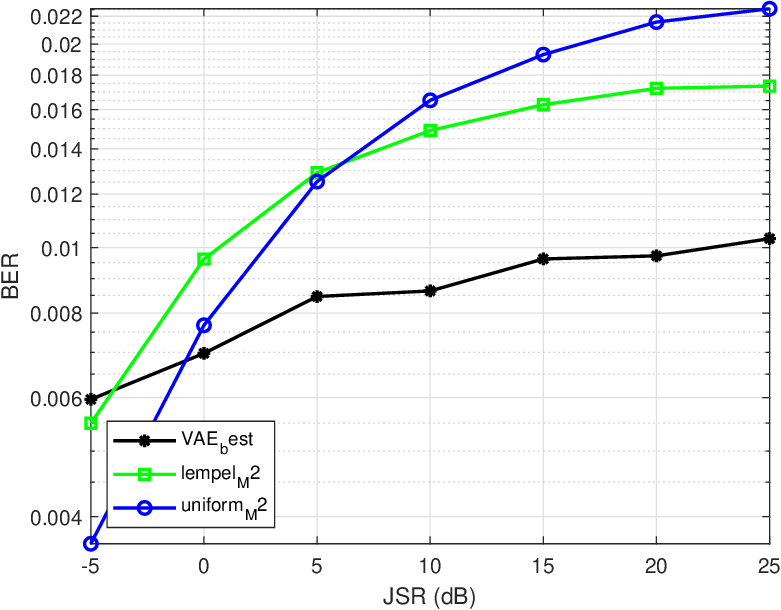}}\hfill
\subcaptionbox{Multi-tone jamming ($2$ tones)\label{fig:vae_mt}}{%
   \includegraphics[width=0.33\textwidth]{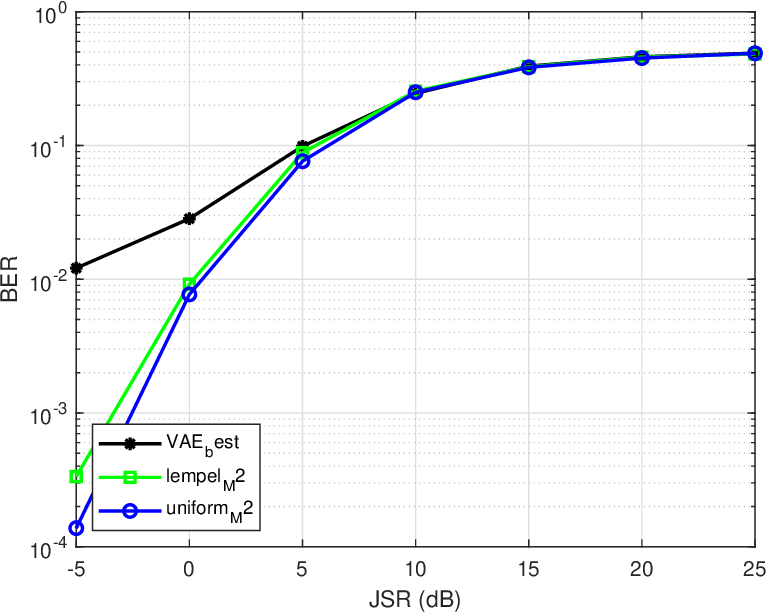}}\hfill
\subcaptionbox{ETU + $200$~Hz Doppler\label{fig:vae_dop}}{%
   \includegraphics[width=0.33\textwidth]{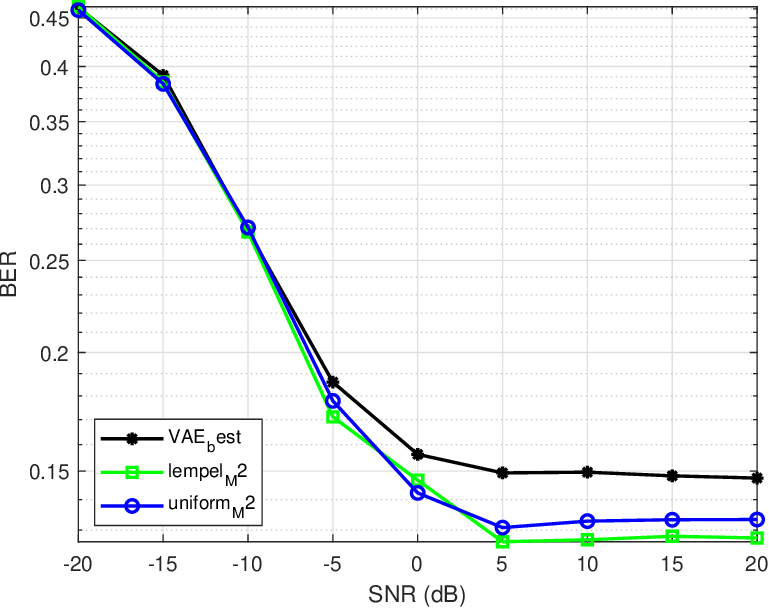}}
\caption{Linear-VAE generator vs.\ Lempel-Costas $M\!=\!2$ and uniform $M\!=\!2$ under (a) pulse,
(b) multi-tone, and (c) ETU $200$~Hz Doppler jamming.}
\label{fig:vae_comparison}
\end{figure*}
\else
A head-to-head comparison of the VAE-best code against Lempel-Costas $M\!=\!2$ and uniform
$M\!=\!2$ across pulse, multi-tone, and ETU $200$~Hz Doppler jamming (at $1000$ frames per
point) is provided in the supplementary material.
\fi

The interpretation is that generative ML, by learning the structural manifold of valid
sparse codes, is able to navigate around the noise that defeats the predictive
pipeline. The trade-off is that the generative result is non-deterministic
(dependent on the random latent sample); for deployment use cases requiring
reproducibility, Lempel-Costas $M\!=\!2$ remains the recommended deterministic choice.

\subsection{Doppler Robustness}\label{sec:res_doppler}

\ifextended
The complement to jamming-axis performance is mobility-axis performance. Under ETU with
$200$~Hz Doppler, the dense regime ($M\!\geq\!180$) achieves the lowest BER ($0.089$ at
SNR$\!=\!10$~dB), the sparse regime ($M\!\leq\!5$) is roughly $20\%$ higher ($0.108$--$0.137$),
and the bad zone ($M\!=\!60$) is intermediate ($0.13$). The mechanism is per-symbol
channel-estimate averaging across the $M$ non-zero positions: under i.i.d.\ Jakes-spectrum
statistics the post-spreading estimate variance scales as $\sigma_{\hat h}^2(M)\!\propto\!1/M$,
predicting the sparse-to-dense improvement. At $500$~Hz, all densities hit the single-pilot-ZF
receiver-saturation floor ($\!\approx\!0.30$), so Doppler beyond approximately $300$~Hz is
receiver-limited rather than code-limited.
\else
On the mobility axis, under ETU with $200$~Hz Doppler the dense regime ($M\!\geq\!180$) achieves
the lowest BER ($0.089$ at SNR$\!=\!10$~dB) while the sparse regime is $\sim\!20\%$ higher; at
$500$~Hz all densities hit the single-pilot-ZF receiver-saturation floor ($\!\approx\!0.30$), so
Doppler beyond $\sim\!300$~Hz is receiver-limited rather than code-limited. The per-column
averaging mechanism ($\sigma_{\hat h}^2(M)\!\propto\!1/M$) is detailed in the supplementary
material.
\fi

\subsection{Final Comparison and Recommendation}\label{sec:res_final}

Table~\ref{tab:final_summary} summarizes the BER performance of the leading candidate
codes across the four-threat operating matrix.

\begin{table*}
\centering
\caption{Recommended-code BER summary (lower is better). All values at the indicated
operating points with $1000$-frame averaging. Best per scenario column in \textbf{bold}.
\textit{Pulse 10k}: periodic pulse jammer at $10$~kHz, $2\%$ duty.
\textit{Multi-tone}: $2$-tone jammer with random tone frequencies.
\textit{Doppler 200/500~Hz}: ETU multipath fading channel with maximum Doppler $f_d=200$
and $500$~Hz, respectively.}
\label{tab:final_summary}
\setlength{\tabcolsep}{4pt}
\begin{tabular}{@{}lcccc@{}}
\toprule
                                          & Pulse 10k          & Multi-tone          & Doppler 200~Hz      & Doppler 500~Hz \\
Code                                      & JSR$=$10 dB        & JSR$=$10 dB         & SNR$=$10 dB         & SNR$=$10 dB \\
\midrule
Walsh                                     & 0.33               & 0.29                & 0.40                & 0.51 \\
LDSC-$M30$~\cite{yang2024}                & 0.12               & \textbf{0.19}       & 0.11                & 0.33 \\
LDSC-Free uniform $M\!=\!2$               & 0.018              & 0.21                & 0.12                & 0.31 \\
Lempel--Costas $M\!=\!2$                  & 0.020              & 0.29                & 0.11                & 0.31 \\
VAE-best $M\!=\!2$                        & \textbf{0.011}     & 0.25                & 0.12                & \textbf{0.30} \\
Free-overlap $M\!=\!180$ (dense)          & 0.24               & 0.28                & \textbf{0.089}      & 0.31 \\
\bottomrule
\end{tabular}
\end{table*}
All reported BER differences in Table~\ref{tab:final_summary} exceed the
$95\%$ Wilson confidence intervals at the corresponding operating points,
confirming statistical significance. Although the density-axis sweep
(Fig.~\ref{fig:density_sweep}) shows a random-allocation pulse minimum at
$M\!=\!5$ (BER $0.024$), the structured uniform-$M\!=\!2$ design achieves a
strictly lower pulse-jamming BER ($0.018$) with smaller code storage and
lower despread complexity (column support $K\!\cdot\!2$ non-zeros vs
$K\!\cdot\!5$ at $M\!=\!5$), motivating $M\!=\!2$ as the recommendation
over $M\!=\!5$.
The recommended design is uniform-spacing $M\!=\!2$ on disjoint supports (or, for
deterministic deployment, Lempel-Costas $M\!=\!2$): both achieve a $\sim\!5\!\times$
reduction in pulse-jamming BER compared to the LDSC-$M\!=\!30$ baseline of~\cite{yang2024}
while keeping multi-tone and Doppler BER within $\sim\!20\%$ of the dense optimum. The
generative-ML candidate (VAE-best) is even better for pulse jamming but lacks
reproducibility; we recommend it as a complementary option when training data is
available and reproducibility is not required.

\ifextended
\textit{Implementation complexity.}
Beyond BER, deployment also depends on code-generation, storage, and
receiver-side complexity. Table~\ref{tab:complexity} summarizes these for
the leading candidate designs. The sparse-$M\!=\!2$ designs (Free uniform
and Lempel--Costas) are uniformly cheaper than the prior-art $M\!=\!30$
Alamouti construction across all three axes: code generation drops from
$O(KM)\!=\!O(L)$ (one Alamouti seed per block, $K/2$ blocks of $M$ vertical
stacks) to $O(K)$ (one disjoint $M$-position assignment per column), per-code
storage drops from $O(KM)$ to $O(K)$, and the matched-filter despreader
shrinks from $O(KM)$ to $O(K)$ multiply-accumulates per symbol. The
Lempel--Costas variant adds $O(\log p)$ for the primitive-root construction
but avoids the random row-permutation needed by the Alamouti family. These
gains are independent of the BER advantages and reinforce the practical
case for sparse density selection.

\begin{table}[!t]
\centering
\caption{Implementation complexity of candidate spreading designs (per
symbol). $L\!=\!360$, $K\!=\!12$; $M$ is the column density. Generation
is per-code, executed once at provisioning.}
\label{tab:complexity}
\setlength{\tabcolsep}{4pt}
\footnotesize
\begin{tabular}{@{}lccc@{}}
\toprule
Design                                    & Generation        & Storage      & Despread \\
\midrule
Walsh / OVSF                              & $O(L\log L)$      & $O(KL)$      & $O(KL)$   \\
LDSC-$M30$~\cite{yang2024}                & $O(KM)\!=\!O(L)$  & $O(KM)$      & $O(KM)$   \\
LDSC-Free uniform $M\!=\!2$               & $O(K)$            & $O(K)$       & $O(K)$    \\
Lempel--Costas $M\!=\!2$                  & $O(K\!+\!\log p)$ & $O(K)$       & $O(K)$    \\
VAE-best $M\!=\!2$                        & training-bound    & $O(K)$       & $O(K)$    \\
Free-overlap $M\!=\!180$ (dense)          & $O(KM)$           & $O(KM)$      & $O(KM)$   \\
\bottomrule
\end{tabular}
\end{table}
\else
\textit{Implementation complexity.} The sparse-$M\!=\!2$ designs (Free uniform and
Lempel--Costas) are uniformly cheaper than the prior-art $M\!=\!30$ Alamouti construction on
all three axes: code generation, per-code storage, and matched-filter despreading each drop
from $O(KM)$ to $O(K)$ per symbol (Lempel--Costas adds $O(\log p)$ for the primitive-root
construction). A full per-design complexity table is in the supplementary material.
\fi


\section{Conclusion and Future Work}\label{sec:conclusion}

We have presented a systematic study of sparse spreading-code design for anti-jamming
NB-IoT direct-sequence spread spectrum communications. Departing from the prior-art
ML-guided LDSC framework of~\cite{yang2024}, we generalize the design space along a
column-density axis, characterize an empirically-discovered adverse density regime at
$M\!\approx\!L/K$, and propose a Free family of codes (disjoint per-column supports
with arbitrary density) that outperforms the density-matched Alamouti-paired construction by
approximately $20\%$ in mean BER across the pulse and multi-tone scenarios. We further demonstrate that
deterministic algebraic constructions (Lempel-Costas) match the empirical optimum, while
the closely related Paley-QR family degrades to the random-decision floor under periodic pulse jamming.
On the ML axis, predictive wBER regression fails on a 210-sample dataset, but generative
modeling (linear VAE) successfully synthesizes jamming-robust codes that beat the best
algebraic construction by $45\%$. We derive and validate three closed-form BER expressions
incorporating a corrective effective processing gain $G_{\rm eff} = L/(2K)$ that accounts
for the composite-power SNR normalization of the $K$-stream waveform and pilot-aided ZF noise.

The recommended design (uniform-spacing $M\!=\!2$ on disjoint supports for
empirical-optimum performance, or Lempel-Costas $M\!=\!2$ for a deterministic
closed-form alternative) achieves a $5\!\times$ reduction in pulse-jamming BER relative
to the prior-art LDSC-$M30$ baseline. More broadly, our results suggest that
future anti-jamming sparse spreading systems should prioritize density
selection before pursuing machine-learning-based coefficient optimization,
since operating near the overlap transition can dominate performance
regardless of optimization strategy.

Future work directions, in order of practical impact:
\ifextended
\begin{enumerate}
\item \textit{Multi-pilot or data-aided channel tracking} to unlock the dense-code
mobility regime beyond the current $\sim\!300$~Hz Doppler ceiling. A $Z$-pilot
generalization of the effective processing gain follows the form
$G_{\rm eff}(Z) = L/[K(1+1/Z)]$, which reduces to our single-pilot
$L/(2K)$ at $Z=1$ and recovers the classical $L/K$ in the limit $Z\to\infty$.
The underlying $Z$-dependent pilot-aided ZF noise penalty $(1+1/Z)\sigma^2$ is
already derived for general $Z$ in Section~\ref{sec:an_pilot}; the remaining
throughput/coverage trade-off analysis (optimal $Z$ and pilot placement) is in
preparation.
\item \textit{Joint position-and-value optimization} via gradient-based or
semi-definite-programming methods, particularly for the algebraic constructions.
\item \textit{Specialized algebraic constructions} (Singer difference sets, perfect
difference sets, Reed--Muller thinning) tailored to specific jammer-type vulnerabilities.
\item \textit{Adversarial / smart-jammer scenarios} where the jammer has knowledge of the
code (sweep, reactive, follower, DRFM-based jammers); a game-theoretic
minimax design framework against this class. The closed-form $G_{\rm eff}$
and sparse pulse-jamming expressions provide the analytical baseline against
which adaptive-jammer strategies can be benchmarked.
\item \textit{Generative ML expansion}: nonlinear VAE/GAN, conditional models trained on
specific jammer-class datasets, reinforcement-learning-driven code synthesis.
\item \textit{Unequal-power / near--far NOMA/SCMA}: relaxing the equal-power assumption of
Section~\ref{sec:res_setup} to power-profile-dependent operation, where the interference at
column $k$ becomes the power-weighted $\sum_{j\neq k}P_j\,|\bc_j^H\bc_k|^2$ and near--far
disparities dominate the cross-correlation impact; joint power-control and sparse-code
co-design for dense SCMA deployments.
\item \textit{Real-channel measurements} for V2X, UAV, and satellite links instead of
statistical LTE models.
\item \textit{Field testing / over-the-air validation} of the recommended design on
hardware.
\end{enumerate}
\else
\begin{enumerate}
\item \textit{Multi-pilot or data-aided channel tracking} to unlock the dense-code mobility
regime beyond the current $\sim\!300$~Hz Doppler ceiling. The $Z$-pilot effective gain
$G_{\rm eff}(Z)=L/[K(1+1/Z)]$ reduces to our single-pilot $L/(2K)$ at $Z=1$ and to the classical
$L/K$ as $Z\to\infty$; its $(1+1/Z)\sigma^2$ ZF penalty is derived for general $Z$ in the
supplementary material (Section~S-II.A), with the throughput/coverage trade-off in preparation.
\item \textit{Joint position-and-value optimization} via gradient-based or SDP methods for the
algebraic constructions.
\item \textit{Specialized algebraic constructions} (Singer and perfect difference sets,
Reed--Muller thinning) tailored to specific jammer types.
\item \textit{Adversarial / smart-jammer scenarios} (sweep, reactive, follower, DRFM jammers) and
a game-theoretic minimax design, benchmarked against our closed-form $G_{\rm eff}$ and pulse model.
\item \textit{Generative ML expansion}: nonlinear VAE/GAN, jammer-class-conditional models, and
reinforcement-learning-driven code synthesis.
\item \textit{Unequal-power / near--far NOMA/SCMA}: relaxing the equal-power assumption of
Section~\ref{sec:res_setup} to power-weighted interference and joint power-control/sparse-code
co-design.
\item \textit{Real-channel measurements} (V2X, UAV, satellite) and \textit{over-the-air
validation} of the recommended design on hardware.
\end{enumerate}
\fi


\setlength{\IEEEbibitemsep}{0pt plus 0.2pt}
\bibliographystyle{IEEEtran}
\bibliography{references}

\end{document}